\documentclass[useAMS,usenatbib]{mn2e}
\usepackage{epsfig}
\usepackage{float}
\usepackage{placeins}
\usepackage{graphicx}
\usepackage{epsfig}
\usepackage{bm}% bold math
\usepackage{amsfonts}
\usepackage{amssymb}
\usepackage{times}
\usepackage{natbib}
\usepackage{color}
\usepackage[normalem]{ulem}

\title[Flux and spectral variations in 3C 279]{Multi-band optical variability of 3C 279 on diverse timescales}
\author[Agarwal et al.]
{Aditi Agarwal$^{1}$\thanks{E-mail: aditiagarwal.phy@gmail.com},
Sergio A. Cellone$^{2,3}$,
Ileana Andruchow$^{4}$,
Luis Mammana$^{3}$,
\newauthor Mridweeka Singh$^{5}$,
G. C. Anupama$^{1}$,
B. Mihov$^{6}$,
Ashish Raj$^{1}$,
L. Slavcheva-Mihova$^{6}$,
\newauthor Aykut \"Ozd\"onmez$^{7}$,
Erg\"un Ege$^{8}$ \\
\\
$^{1}$Indian Institute of Astrophysics, Block II, Koramangala, Bangalore, India, 560034\\
$^{2}$Facultad de Ciencias Astron\'omicas y Geof\'isicas, Universidad Nacional de La Plata, Paseo del Bosque, B1900FWA, La Plata, Argentina \\
$^{3}$Complejo Astron\'omico "El Leoncito" (CASLEO), CONICET-UNLP-UNC-UNSJ, San Juan, Argentina  \\
$^{4}$Instituto de Astrof\'isica de La Plata (CCT La Plata - CONICET - UNLP), La Plata, Argentina \\
$^{5}$Aryabhatta Research Institute of Observational Sciences (ARIES),
Manora Peak, Nainital -- 263002, India\\
$^{6}$  Institute of Astronomy and NAO, Bulgarian Academy of Sciences,72 Tsarigradsko Chaussee Blvd., 1784 Sofia, Bulgaria \\
$^{7}$ T$\ddot{U}$B$\dot{I}$TAK National Observatory , 07058, Akdeniz University Campus, Antalya, Turkey\\
$^{8}$Istanbul University, Graduate School of Science and Engineering, 34116, Beyazıt, Istanbul, Turkey \\
}

\begin{document}
\newdimen\digitwidth
\setbox0=\hbox{2}
\digitwidth=\wd0
\catcode `#=\active
\def#{\kern\digitwidth}

\date{Accepted ....... Received  ......; in original form ......}

\pagerange{\pageref{firstpage}--\pageref{lastpage}} \pubyear{2019}

\maketitle

\label{firstpage}

\begin{abstract}

We have monitored the flat spectrum radio quasar, 3C 279, in the optical $B$, $V$, $R$ and $I$ passbands
from 2018 February to 2018 July for 24 nights, with a total of 716 frames,
to study flux, colour and spectral variability on diverse timescales. 3C\,279 was
observed using seven different telescopes: two in India, two in Argentina, two in Bulgaria and one in Turkey to understand the nature of the source in optical regime.
The source was found to be active during the whole monitoring period and displayed significant flux variations in $B$, $V$, $R$, and $I$ passbands.
Variability amplitudes on intraday basis varied from 5.20\% to 17.9\%.
A close inspection of variability patterns during our observation cycle reveals simultaneity among optical emissions from all passbands.
During the complete monitoring period, progressive increase in the amplitude of variability with frequency was detected for our target.
The amplitudes of variability in $B$, $V$, $R$ and $I$ passbands have been estimated to be 177\%, 172\%, 171\% and 158\%, respectively.
Using the structure function technique, we found intraday timescales ranging from $\sim 23$ minutes to about 115 minutes. We also studied colour-magnitude relationship and found indications of mild
bluer-when-brighter trend on shorter timescales.
Spectral indices ranged from 2.3 to 3.0 with no clear trend on long term basis.
We have also generated spectral energy distributions for 3C\,279 in optical $B$, $V$, $R$ and $I$ passbands for 17 nights.
Finally, possible emission mechanisms causing variability in blazars are discussed briefly.

 \end{abstract}
 

\section{Introduction}
\label{sec:Introduction}

The term Active Galactic Nuclei (AGN) is used to describe small bright regions in the center of certain galaxies
with characteristic bolometric luminosities ranging between 10$^{41}$--10$^{48}$ erg s$^{-1}$.
AGNs are believed to be powered by an actively accreting and possibly spinning central
super massive black hole (SMBH) (Begelman et al., 1984).
AGNs are known to host SMBHs with masses ranging from $10^6~M_{\odot}$ -- $10^{10}~M_{\odot}$.
The central SMBH accretes matter due to its strong gravitational forces and surrounding it is an optically thick, geometrically thin
accretion disc (AD) formed due to the loss of angular momentum through viscous and turbulent processes coming into
play during accretion (Shakura \& Sunyaev 1973). The AD emits mainly in optical, UV and soft X-ray bands of the electromagnetic (EM) spectrum.
In addition to the central SMBH and the AD, the canonical model of an AGN
consists of a dusty torus surrounding %away from
 the central region,
an X-ray emitting corona, and relativistic bipolar outflows (Blandford \& K{\"o}nigl, 1979).
A special class of AGNs having relativistic jets pointing towards the observer are known as blazars.
Blazars are characterized by strong emission violently variable over
the entire electromagnetic (EM) spectrum and also apparent super-luminal motions.
They are also the dominant sources for gamma-ray emission 
in the sky (Acero et al.\ 2015).
Blazars have been classified into BL Lac objects (BL\,Lacs) and flat spectrum radio quasars (FSRQs)
based on their optical spectra (Giommi et al. 2012). 
A high degree of linear polarization at optical wavelengths has been reported for some blazars, including 3C\,279 (Blinov et al. 2015).
Andruchow et al.\ (2003) detected strong microvariability of $\sim$10\% in the linear polarization of 3C 279 observed in V passband.
Blazars have their jet axis aligned at angles $\leq 10^{\circ}$ to the observer's line of sight (LOS, e.g., Urry \& Padovani 1995) which
along with the relativistic beaming leads to above observational features of blazars.
It is generally believed that the jet alignment increases the amplitude of the emission and contracts the variability timescale.
Blazars have been known to vary throughout the EM spectrum over diverse timescales ranging from few minutes to years (Fan et al.\ 2005).
Blazars variability has been broadly divided into three classes: magnitude changes of upto few tenths over a time-scale of few minutes to day or less are considered to be intraday variability
(IDV; Wagner \& Witzel 1995, Xie et al.\ 2004) or microvariability, flux changes typically exceeding $\sim$ 1 magnitude over several days to months are known as short term variability (STV), while
the changes over several months to years (sometimes can exceed even $\sim$ 5 magnitudes) are grouped under long term variability (LTV;  Fan \& Lin 2000; Gupta et al.\ 2004).

\begin{table*}
\caption{ Details of telescopes and instruments}
\textwidth=7.0in
\textheight=9.0in
\vspace*{0.2in}
\noindent
\begin{tabular}{p{1.4cm}p{1.6cm}p{1.6cm}p{1.6cm}p{1.6cm}p{1.6cm}p{1.6cm}p{1.6cm}p{1.6cm}} \hline
Site:             &     A                     & B                            &  C                                                      & D                            &E                                   &F                                              & G                               & H                               \\\hline
Telescope         & 2.01-m RC                 & 1.30-m RC Cassegrain         &  2.15 m RC Nasmyth                                      & 2.15~m RC Cassegrain         &  0.6~m classic Cassegrain          & 2-m Ritchey-Chr\'etien + FoReRo-2 focal reducer & 50/70-cm Schmidt                & 0.6m RC Cassegrain\\
CCD model         & SITe ST-002               & Andor $512 \times 512$                &  Roper Scientific Versarray 2048B (cooling: liquid N2)  & TK1024 (cooling: liquid N2)  & SBIG STL-1001E (cooling: Peltier)  &  VersArray:1300B                              &  FLI PL16803                    &  Apogee Alta U42 CCD  \\
Chip size         & $2048 \times 2048$ pixels   &  $512 \times 512$ pixels
&  $2048 \times 2048$ pixels                                & $1024 \times
1024$ pixels      & $1024 \times 1024$ pixels            &  $1340 \times
1300$ pixels                      &  $4096 \times 4096$ pixels        &
$2048 \times 2048$ pixels \\
Pixel size        & $15\times 15~\mu$m        & $16 \times 16~\mu$m
& $13.5 \times 13.5~\mu$m                                   &   $24 \times
24~\mu$m         &  $24 \times 24~\mu$m                & $20.0 \times
20.0~\mu$m                        & $9 \times 9~\mu$m                &
$13.5 \times 13.5~\mu$m \\
Scale             & $0.296\arcsec$/pixel      &$0.63\arcsec$/pixel           & $0.45\arcsec$/pixel                                     &$0.27\arcsec$/pixel (unbinned)& $0.54\arcsec$/pixel (unbinned)     &$0.737\arcsec$/pixel                           &  $1.079\arcsec$/pixel           &     $0.58\arcsec$/pixel        \\
Field             &$10\arcmin\times10\arcmin$ &$5.4\arcmin\times5.4\arcmin$  & 9$\arcmin$ diameter                                     & 5$\arcmin$ diameter          & $9.3\arcmin\times9.3\arcmin$       & $16.5\arcmin\times16\arcmin$                  &  $73.7\arcmin\times73.7\arcmin$ &   $19.8\arcmin\times19.8\arcmin$ \\
Gain              & 1.22~$e^-$/ADU             & 1.4~$e^-$/ADU                 & 2.18~$e^-$/ADU                                           &  1.98~$e^-$/ADU              & 2.2$e^-$/ADU                       &1.0$e^-$/ADU                                   & 1.49~$e^-$/ADU                   &  1.19~$e^-$/ADU             \\
Read Out Noise    & 4.87~$e^-$ rms             & 6.0~$e^-$ rms                 & 3.1~$e^-$ rms                                           &  7.40~$e^-$ rms               &  14.8~$e^-$ rms                     &2.0~$e^-$ rms                                  &8.97~$e^-$ rms                   &   11.0~$e^-$ rms               \\
Binning used      & 1$\times$1                &  1$\times$1                  &   1$\times$1                                            &   2$\times$2                 & 1$\times$1                         &  1$\times$1                                   &  1$\times$1                     & 1$\times$1 \\
Typical seeing    & 1$\arcsec$ to 3$\arcsec$  & 1$\arcsec$ to 3$\arcsec$     & 1.8$\arcsec$ to 2.5$\arcsec$                            & 1.8$\arcsec$ to 2.5$\arcsec$ &1.2$\arcsec$ to 2.2$\arcsec$        & 1.5$\arcsec$ to 2.5$\arcsec$                  &  2.0$\arcsec$ to 3.0$\arcsec$   & 1.5$\arcsec$ to 2.5$\arcsec$  \\\hline
\end{tabular} \\
\noindent
 A    : 2.01-m RC Himalayan Chandra Telescope (HCT) at Indian Astronomical Observatory, Hanle, India \\
 B    : 1.30-m RC Cassegrain optical telescope, ARIES, Nainital, India \\
 C,D  : 2.15-m RC Nasmyth at CASLEO, Argentina \\
 E    :  0.6 m classic Cassegrain at CASLEO, Argentina \\
 F  : 2-m RC telescope at National Astronomical observatory Rozhen, Bulgaria \\
 G  : 50/70-cm Schmidt telescope at National Astronomical Observatory, Rozhen, Bulgaria   \\
 H  : 0.6m RC Cassegrain telescope at Ulupınar observatory operated by Istanbul University, Turkey \\
\end{table*}

The spectral energy distribution (SED) of blazars is characterized
by two broad bumps: a low-energy one due to synchrotron 
radiation, which covers the radio to the X-ray range, and a high-energy one due to inverse Compton (IC) emission, which covers
from the X-rays to the $\gamma$-rays. According to the leptonic jet models, the low-frequency emission can be explained as synchrotron
emission from non-thermal electrons. On the other hand, the high-energy radiation can be associated with the inverse-Compton (IC) scattering of
low-energy synchrotron photons from the jet (synchrotron self-Compton or SSC, \& K{\"o}nigl 1981), and/or with the thermal photons outside the jet (External Compton, EC,  Hunger \& Reimer 2016).
The SEDs of blazars can also be generated
using other models e.g. hadronic or lepto-hadronic emission models (e.g., M{\"u}cke et al. 2003; B{\"o}ttcher et al.\ 2013).
SEDs are helpful in identifying the contributions of emission from synchrotron mechanism, dust, broad line region (BLR), AD, star light and surrounding regions.

3C\,279 is a well studied blazar (e.g. Maraschi et al. 1994; Wehrle et al. 1998; Lindfors et al. 2006;
Collmar et al. 2007) which has shown multiwavelength flux variability. It is the first extragalactic radio source which showed
superluminal motion (Cohen et al. 1971). 3C 279 is a luminous FSRQ at $z = 0.536$ (Lynds et al. 1965) with the central black hole mass
in the range of $(3\mbox{--}8)\times {10}^{8}\,{M}_{\odot }$ (Gu et al. 2001; Woo \& Urry 2002; Nilsson et al. 2009).
It is also the first FSRQ detected in VHE $\gamma$-rays by the Major Atmospheric Gamma-ray Imaging Cherenkov (MAGIC) telescope
(MAGIC Collaboration 2008). 
Extensive monitoring with
high-resolution VLBI observations revealed many important results for this source, e.g. a one-sided jet
extending south-west on pc-scale characterized by bright knots ejected from the core region, etc. (see Unwin et al. 1989;
Wehrle et al. 2001; Jorstad et al. 2005; Chatterjee et al. 2008). The electric field vector was found to be aligned with the
jet direction in VLBA polarimetry observations, which indicate that the magnetic field was predominantly perpendicular to the
relativistic jets. The important parameters, e.g. bulk Lorentz factor, viewing angle of the jet flow, were
estimated as $\Gamma_{\mathrm{j}} = 13.3 \pm 0.6$ and $\Theta_{\mathrm{j}} = 1\fdg9 \pm 0\fdg6$ from
VLBA radio observations (Jorstad et al.\ 2017).

Blazar variability studies can provide information on the dominant emission mechanism behind observed manifestations which in turn can shed light on various theoretical models.
Variability studies of blazars have been conducted by optical astronomers around the world for over 50 years but some pertinent questions still remain unresolved.
To further understand the characteristics of 3C\,279, we here study its variability properties on diverse timescales in optical $B$, $V$, $R$, and $I$ bands. 
The photometric data have been obtained from
seven different optical telescopes around the world during 2018 in $B$, $V$, $R$, and $I$
passbands. We have also investigated spectral energy distributions (SEDs) of
the target in the optical regime.
This paper is structured as follows: Section 2 gives an overview of the multiband observations used and data reduction procedure, in Section 3 various analysis techniques are introduced.
Section 4 gives results while discussion is given in Section 5 and conclusions in Section 6.

\section{Observations and Data Reduction}

For this study, observations covering optical $B$, $V$, $R$, and $I$ passbands were performed for the blazar 3C\,279, from 2018 February $-$ 2018 July.
Our photometric observations were performed using seven optical telescopes around the world which are briefly described below. 

We carried our optical $B$, $V$, $R$, and $I$ observations using the 2.01-m optical-infrared Himalayan Chandra Telescope (HCT; telescope A of Table 1) located at Indian Astronomical Observatory, Hanle
(latitude $32^{\circ}$ $46'$\,N, longitude $78^{\circ}$ $57'$\,E, altitude 4500\,m), India, remotely operated from Centre for Research and Education in Science \& Technology (CREST), Hosakote, via a dedicated
satellite link. It has a Ritchey-Chr\'etien (RC) optics with an altitude over azimuth mount. Observations were performed with the Hanle Faint Object Spectrograph Camera (HFOSC) mounted on HCT and equipped
with 2k$\times$4k SITe ST-002 CCD. The central region of 2k$\times$2k with a plate scale of 0.296 arcsec/pixel corresponds to a field of $10\arcmin\times10\arcmin$. More details are given in Table 1.
We have also used 1.3-m Devasthal fast optical telescope (DFOT) of ARIES (telescope B of Table 1), Nainital with latitude $29^{\circ}$ $21'$\,N, longitude $79^{\circ}$ $41'$\,E, altitude 2420\,m
operated by ARIES, Nainital, India. The 1.3-m DFOT has fork equatorial type
mount system and a fast beam with a focal ratio f/4 which provides 40$^{\prime\prime}$ sky view in 1 mm scale at the focal plane. It is
equipped with Andor 2K $\times$ 2K CCD with $13.5~\mu$m pixel size, 512 $\times$ 512 CCD with 16 $\mu$m pixel size, and also a
3326 $\times$ 2504 CCD with $5.4~\mu$m pixel size. For our observations, we have used $512 \times 512$ CCD.
Sky brightness as measured on a moonless night in the $V$ passband is $\sim 21.2$~mag/arcsec$^2$ which varies with the Moon's phase.
It uses RC Cassegrain design and has a field of view of 5.4$^{\prime}$. In
addition to the above two telescopes, optical observations of 3C\,279 were also obtained with two different telescopes
at CASLEO (Argentina): the Helen Sawyer Hogg (HSH; telescope E of Table 1) 0.6\,m telescope (on loan
from the University of Toronto, Canada), and the Jorge Sahade (JS; telescope C,D of Table 1) 2.15\,m
telescope. HSH is equipped with a SBIG STL-1001E CCD camera, while two
different Nitrogen-cooled CCDs were used at the JS: a Tektronix TK1024, and
a Roper Versarray 2048B. Standard Johnson ($BV$) $-$ Cousins ($RI$) filter sets
were used at both telescopes. Furthermore, observations of 3C\,279 were also obtained using the 2-m $f/8$ RC (telescope F of Table 1)
and the 50/70-cm $f/3.44$ Schmidt (telescope G of Table 1) telescopes of the Rozhen National
Astronomical Observatory, Bulgaria, during the period 2018 March 15 to 2018 May 14.
The two-channel focal reducer FoReRo-2 (Jockers et al. 2000) was attached to the
RC focus of the 2-m telescope which yields a focal ratio of $f/2.8$.
Both telescopes are equipped with CCD cameras and a standard Johnson-Cousins $BVRI$
set of filters. Each 3C\,279 observing set consists of several exposures through the
$R$ filter (the 2-m telescope data) or $BVRI$ filters (the Schmidt telescope data;
only the central $1024\times 1024$ pixels of the CCD were actually used). 
Observation of 3C\,279 on 2018 May 13 $-$ 14 made use of the 0.6m telescope (IST60; telescope H of Table 1) at Ulupınar observatory operated by Istanbul University in Turkey.
The telescope is equipped with an Apogee CCD detector and Bessel $UBVRI$ filters. Further details of telescopes used are given in table 1.

CCD images of our source obtained using above telescopes are raw frames which could be affected and deformed by the atmospheric effects or
bad focusing. So, to extract useful information they were subjected to pre-processing, processing and post-processing.
The CCD images obtained with the above mentioned telescopes were de-biased and flat-fielded using standard procedures.
 Finally, all source images were corrected for cosmic rays.
Above steps were performed using IRAF\footnote{IRAF is distributed by the National Optical Astronomy Observatories, which are operated
by the Association of Universities for Research in Astronomy, Inc., under cooperative agreement with the
National Science Foundation.} software. 
The next stage in data reduction is processing which includes extraction of target's position and magnitude from the
rectified CCD intensity array using the Dominion Astronomical
Observatory Photometry (DAOPHOT II) software (Stetson 1987; Stetson 1992). Aperture photometry was performed using DAOPHOT II
for four aperture radii, i.e., $\sim 1 \times$~FWHM, $2 \times$~FWHM, $3 \times$~FWHM and $4 \times$~FWHM,
out of which aperture $2 \times$~FWHM was selected to get instrumental magnitude of the source, as it showed the best S/N ratio.
In addition to above two softwares, MATLAB was used to write any additional program used in data analysis.

For our source 3C\,279, we selected three local standard stars from the observed frame.
Instrumental magnitudes of the target plus these standard stars were extracted using the process described above.
Of these three stars, we finally selected two standard stars with magnitudes
similar to our source and also in its close proximity.
Since we have selected the target and the standard star from the same field,
the air mass along with the instrumental and weather conditions are
the same, making the flux ratios very reliable. The complete observation log is given in Table 2 where column 1 gives the observation date, column 2 reports the telescope used, while column 3
states the number of frames observed in each frame for the respective observation date.
On almost every observation date, we took quasi-simultaneous single data points in $B$, $V$, $R$, and $I$ filters.

\begin{table}
\caption{Log of photometric observations of 3C\,279.}
\textwidth=8.0in
\textheight=11.0in
 \centering
\vspace*{0.2in}
\noindent

\begin{tabular}{lclllll} \\
\multicolumn{2}{|c|}{}\\\hline

  Date of        & Telescope      & \multicolumn{4}{c}{Number of data points}  \\
observations    &     &     \\
(yyyy mm dd)     &  & $B$ & $V$ & $R$ & $I$      \\\hline

2018 02 08    & C  &0 & 0 & 36 & 0  \\
  2018 02 17    & A  &1 & 1 & 47 & 1  \\
   2018 02 18    & A  &1 & 1 & 1 & 1  \\
  2018 02 21   & A &1 & 1 & 57 & 1  \\
  2018 03 15  & G  &3 & 1 & 3 & 2 \\
   2018 03 26  & E  &4 & 4 & 5 & 6 \\
   2018 03 28  & D  &2 & 4 & 4 & 3 \\
  2018 04 16  & B  &2 & 2 & 49 & 2 \\
  2018 04 17  & B &1 & 1 & 53 & 1 \\
   2018 04 18  & B &0 & 0 & 102 & 0 \\
  2018 04 20  & E  &0 & 7 & 7 & 7 \\
  2018 04 21  & E  &0 & 8 & 8 & 8 \\
  2018 04 21  & F  &0 & 0 & 3 & 0 \\
  2018 04 22  & F  &0 & 0 & 3 & 0 \\
   2018 04 23  & G  &1 & 2 & 2 & 2 \\
  2018 04 23  & E  &0 & 3 & 2 & 2 \\
  2018 05 05    & B  &1 & 2 & 86 & 2  \\
  2018 05 06    & B  &2 & 2 & 80 & 2  \\
  2018 05 13    & H  &0 & 0 & 23 & 17  \\
   2018 05 13    & G  &1 & 1 & 2 & 2  \\
   2018 05 14    & F  &0 & 0 & 3 & 0  \\
  2018 05 18    & A  &1 & 1 & 1 & 1  \\
  2018 05 19    & A  &1 & 1 & 1 & 1  \\
  2018 07 08    & D  &0 & 6 & 4 & 3  \\
  \hline
       \end{tabular} 

\end{table}

\section{Analysis Techniques}

To quantify optical variability in the light-curves (LCs) of the source we have employed
three statistics (e.g., de Diego 2010), namely the $C$, $F$, and $\chi^{2}$ tests.
Given the modest number of observations, we opted to use the $C$, $F$, and $\chi^{2}$ tests.

\subsection{$C$-Test}

The C-statistic, introduced by Romero, Cellone, \& Combi (1999), is the most frequently used criterion to claim the variability of the source. The variability detection parameter, C, is defined as the
average of $C_{1}$ and $C_{2}$ with

\begin{equation}
C_{1} = \frac {\sigma(BL-S_{A})}{\sigma(S_{A}-S_{B})}~
,~
C_{2} = \frac {\sigma(BL-S_{B})}{\sigma(S_{A}-S_{B})}.
\end{equation}

Here (BL$-$S$_{A}$), (BL$-$S$_{B}$), and (S$_{A}$$-$S$_{B}$) are the differential instrumental magnitudes of 
the blazar and standard star A (S$_{A}$), the blazar and standard star B (S$_{B}$), and S$_{A}$ vs.\ S$_{B}$
calculated using aperture photometry of the source and comparison stars, while $\sigma$(BL$-$S$_{A}$), 
$\sigma$(BL$-$S$_{B}$) and $\sigma$(S$_{A}$$-$S$_{B}$) are observational scatters of the differential instrumental 
magnitudes of the blazar$-$S$_{A}$, blazar$-$S$_{B}$, and S$_{A}$$-$S$_{B}$, respectively.
Zibecchi et al.\ (2017) analyzed intraday variability in AGNs using different statistical methods currently used in the literature. Through their study they concluded that even though the
C statistics cannot be considered as a genuine statistical test, it could nevertheless be a suitable parameter to detect variability with more reliable results as compared to F-test.

A light curve is considered to be variable at a nominal confidence level of $> 99$\% when $C \geq 2.57$ else we call it non-variable (NV).

\begin{figure*}
\epsfig{figure=  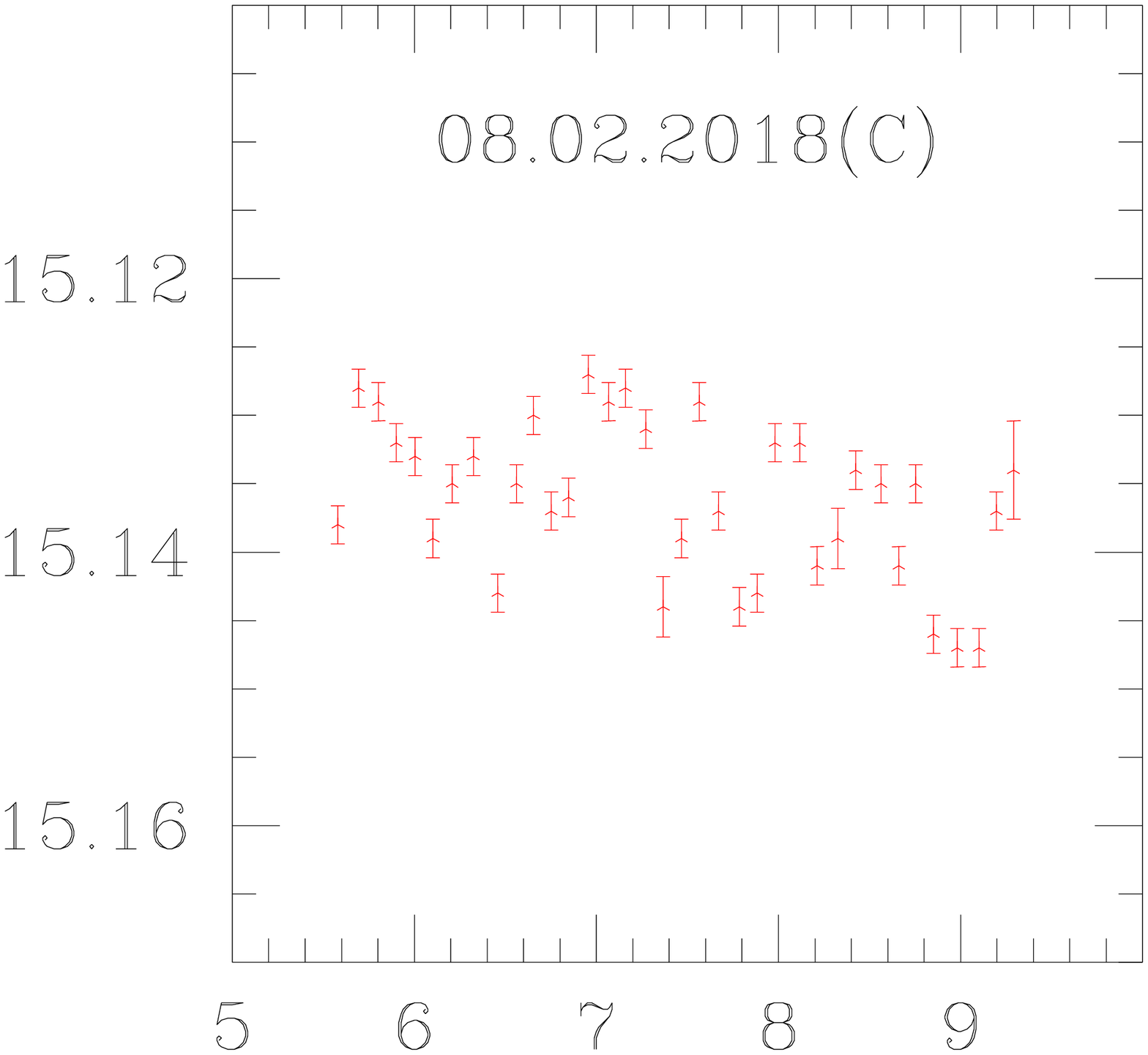,height=2.1in,width=2in,angle=0}
\epsfig{figure=  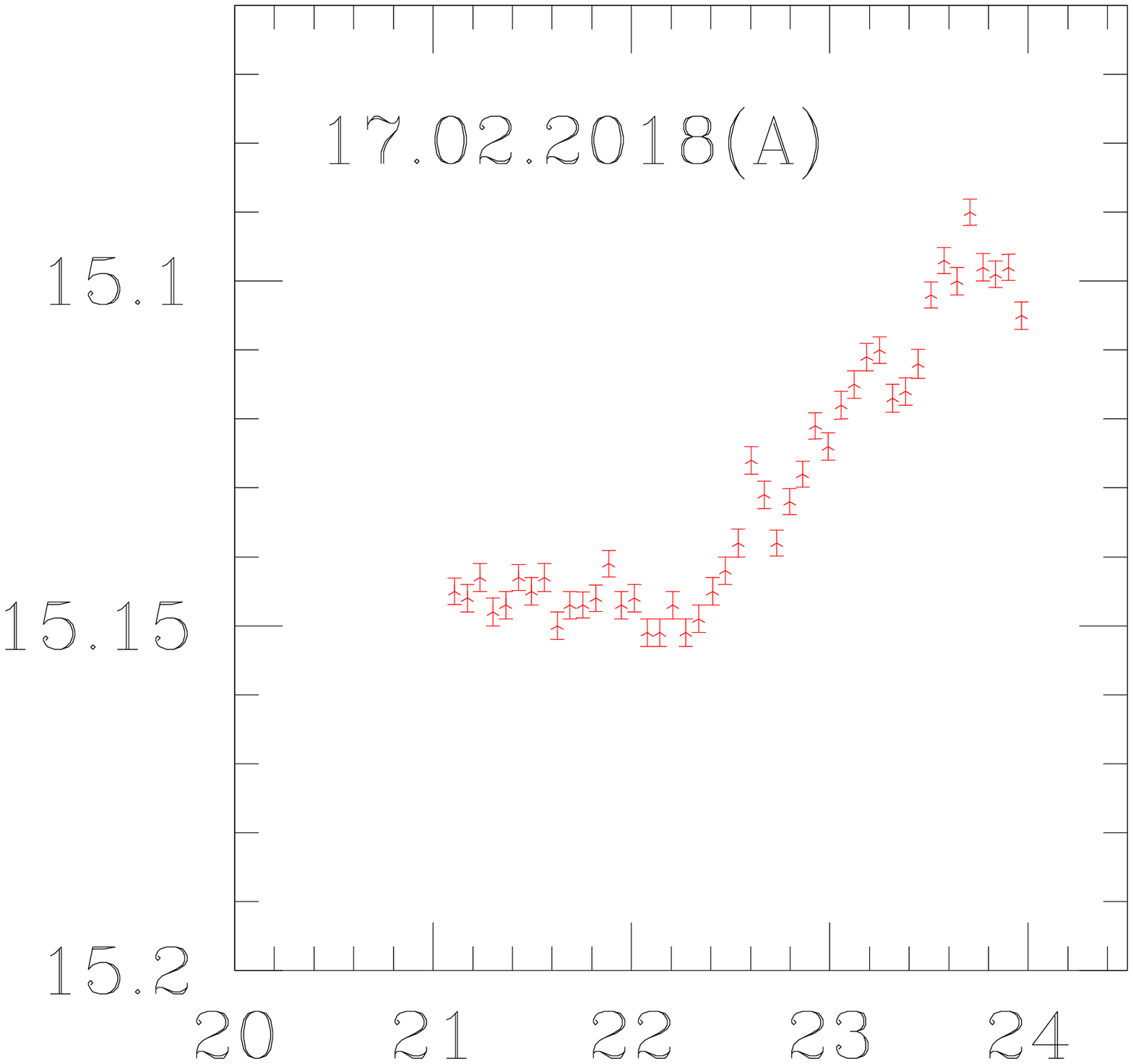,height=2.1in,width=2in,angle=0}
\epsfig{figure=  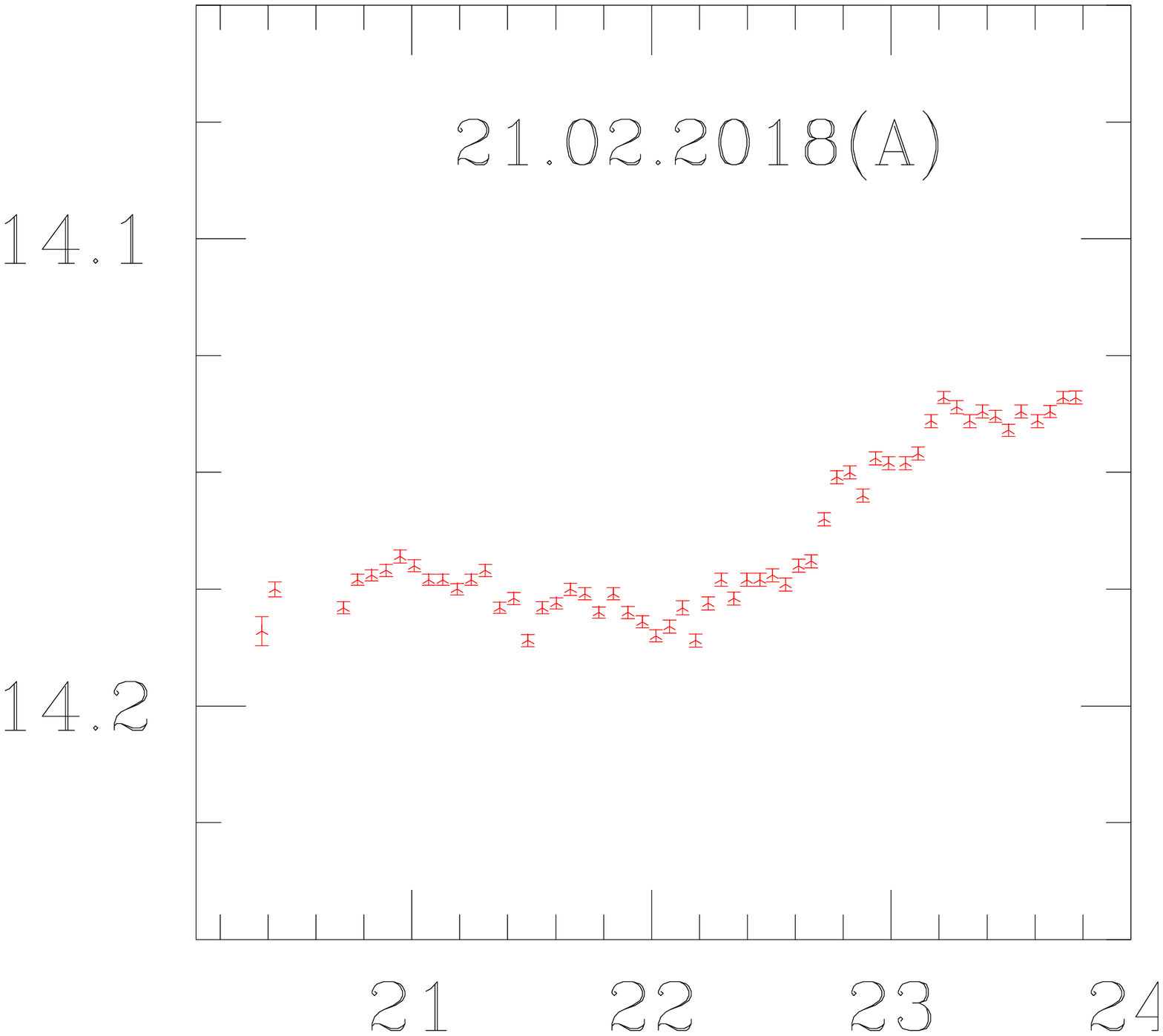,height=2.1in,width=2in,angle=0}
\epsfig{figure=  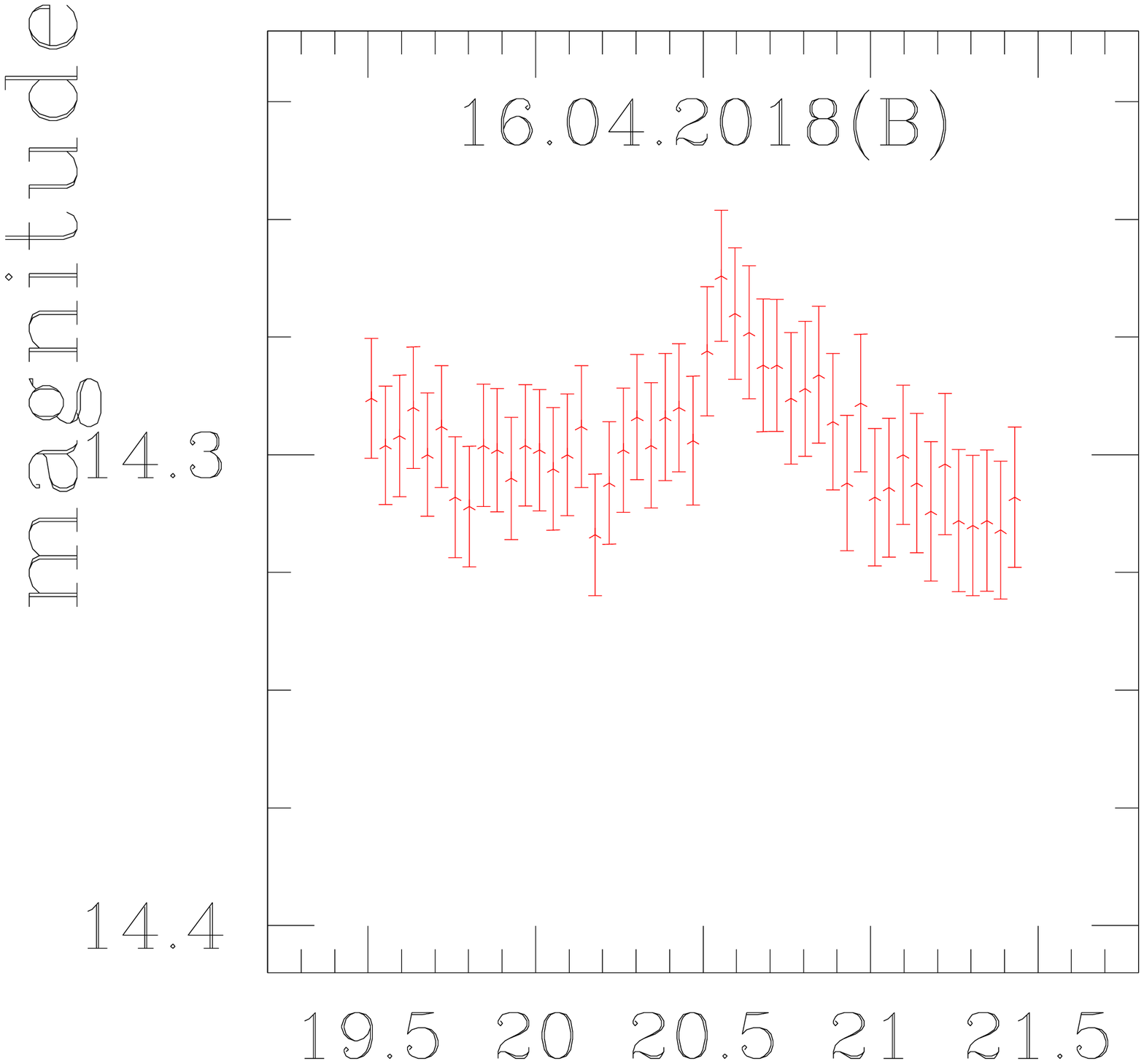,height=2.1in,width=2in,angle=0}
\epsfig{figure=  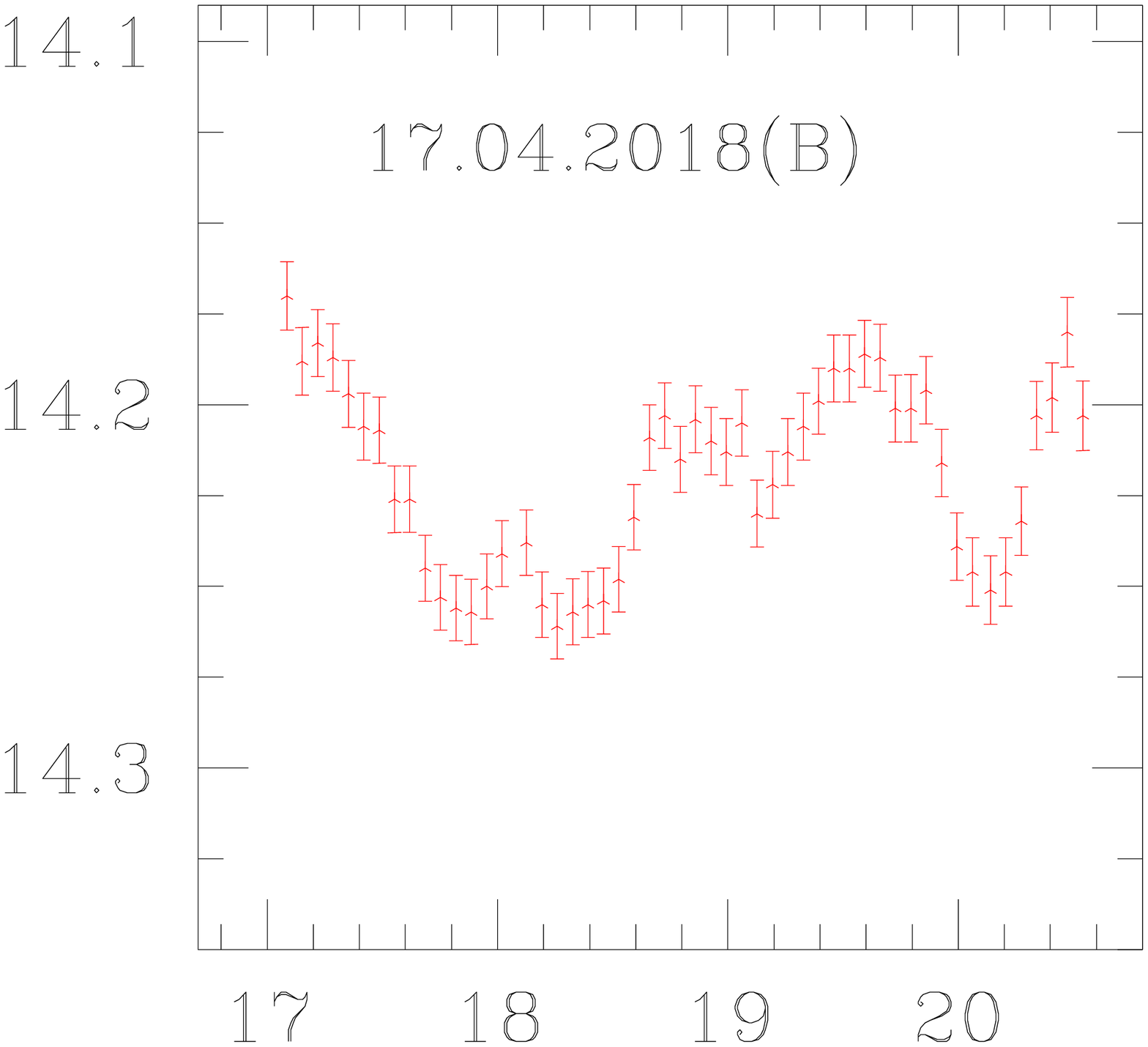,height=2.1in,width=2in,angle=0}
\epsfig{figure=  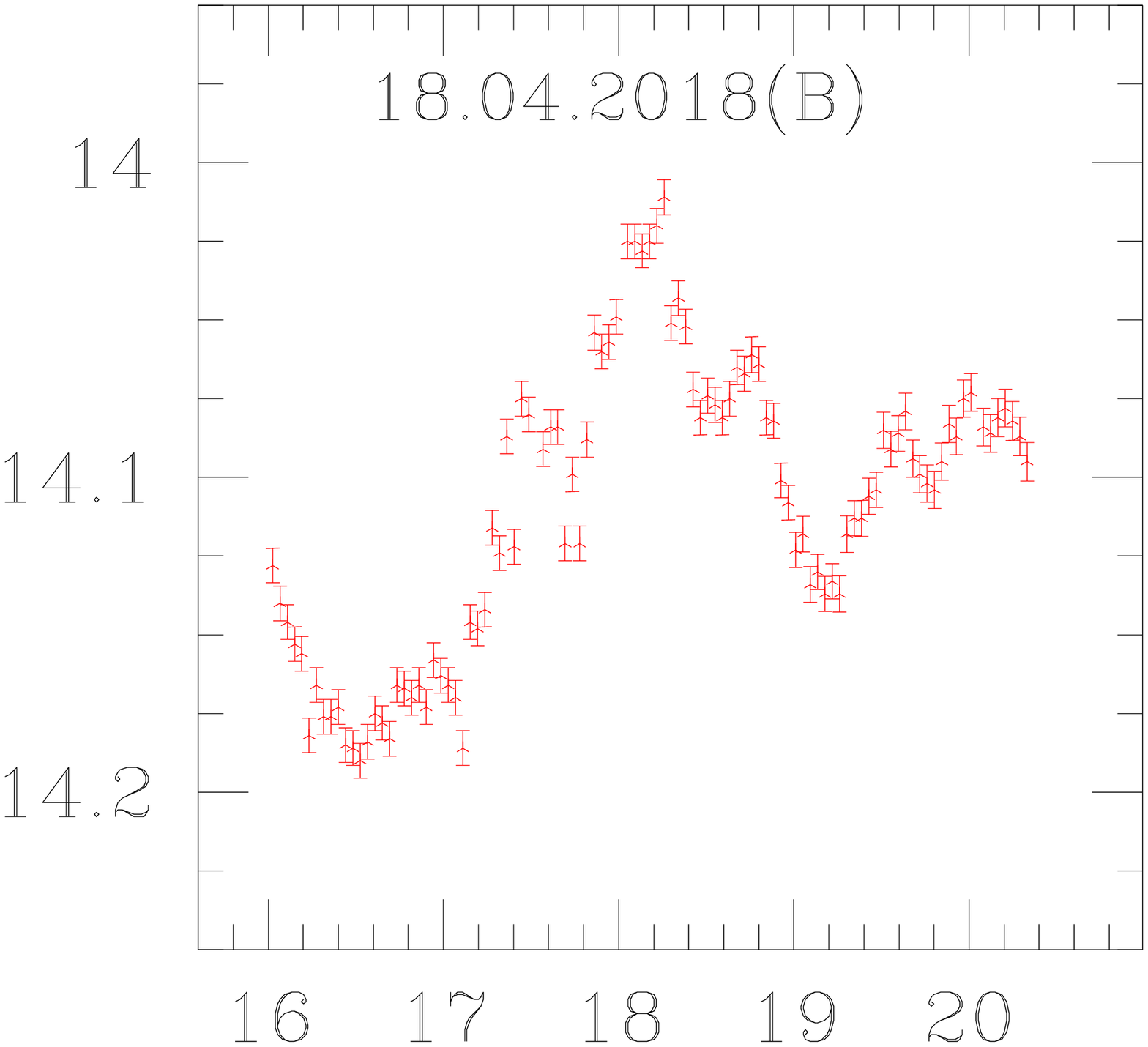,height=2.1in,width=2in,angle=0}
\epsfig{figure=  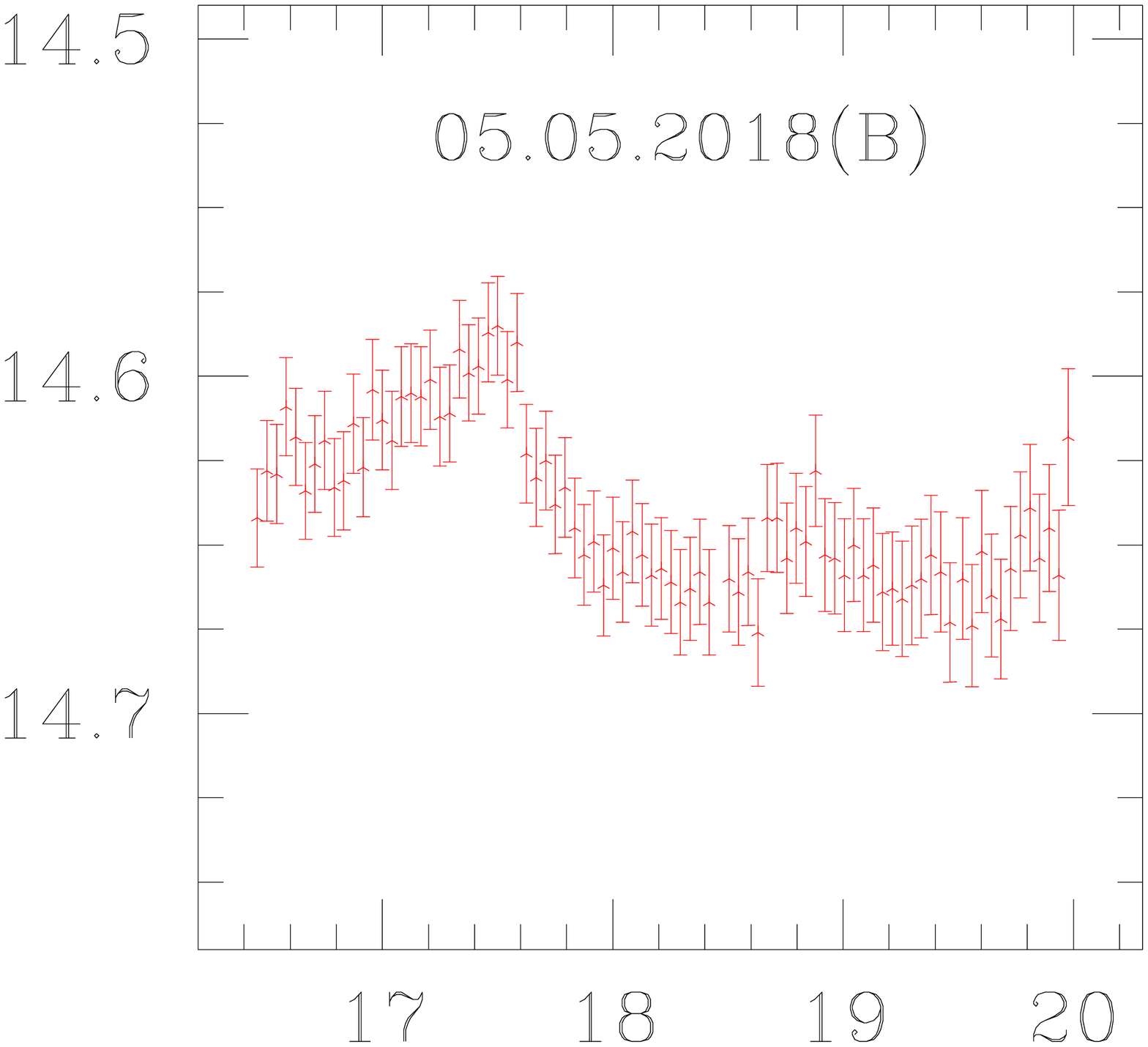,height=2.1in,width=2in,angle=0}
\epsfig{figure=  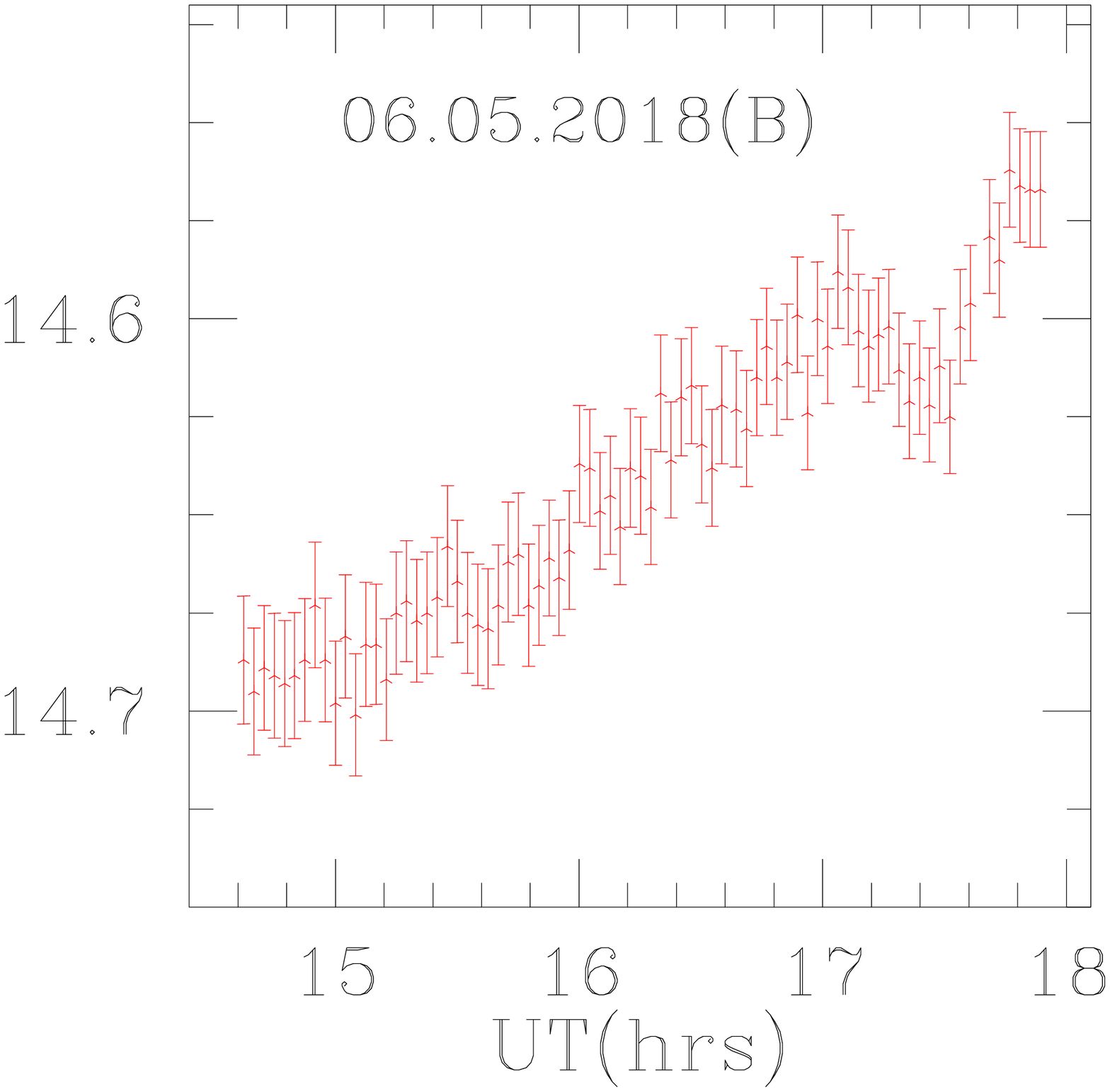,height=2.1in,width=2in,angle=0}
\epsfig{figure=  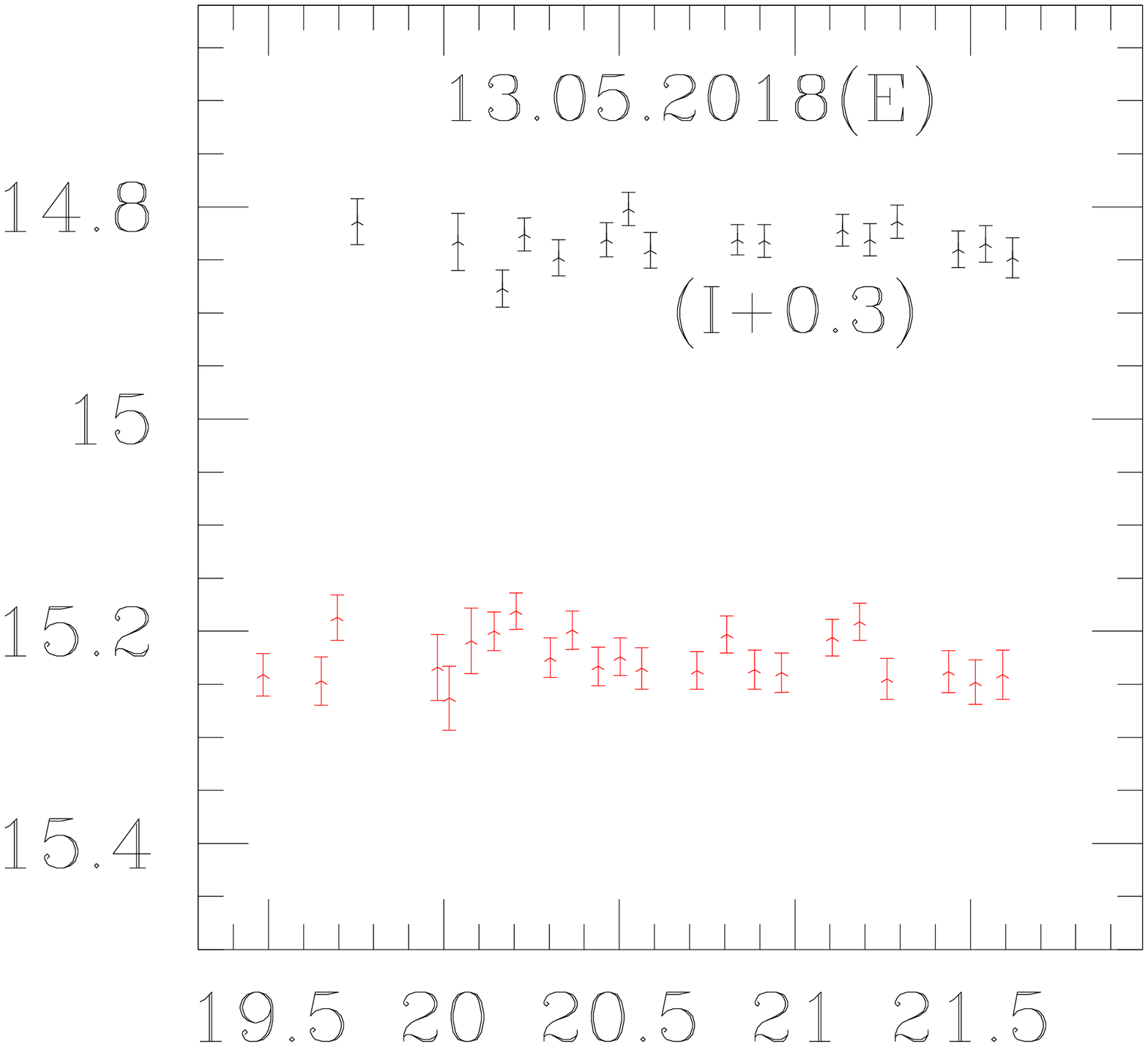,height=2.1in,width=2in,angle=0}
  \caption{Light curves for 3C\,279; red denotes $R$ filter while black denotes $I$ filter. In each plot, X and Y axis are the UT and magnitude, respectively.
Observation date and the telescope used are indicated in each plot.}
\label{LC_BL}
\end{figure*}

\subsection{$F$-Test}

As mentioned by de Diego (2010), the $F$-test is considered to be a powerful and properly distributed statistic, which is used to quantify the variability
nature of sources.
The $F$ values compare two sample variances and are given as:

\begin{equation}
 \label{eq.ftest}
 F_1=\frac{\sigma^2(\mathrm{BL-S_A)}}{\sigma^2(\mathrm{S_A-S_B})}, \nonumber \\ 
 F_2=\frac{\sigma^2(\mathrm{BL-S_B})}{\sigma^2(\mathrm{S_A-S_B})}.
\end{equation}

Here ($\mathrm{BL-S_A}$), ($\mathrm{BL-S_B}$), and ($\mathrm{S_A-S_B}$) are the differential instrumental magnitudes of blazar and standard A, blazar 
and standard B, and standard A and standard B, respectively, computed using the aperture photometry technique, while {$\sigma^2(\mathrm{BL-S_A}$), $\sigma^2(\mathrm{BL-S_B}$), and $\sigma^2(\mathrm{S_A-S_B}$) are 
the variances of differential instrumental magnitudes.
Averaging $F_1$ and $F_2$ gives the mean observational $F$ value which is then compared with the critical value, $F^{(\alpha)}_{\nu_\mathrm{bl},\nu_*}$,
where $\nu_\mathrm{bl}$ and $\nu_*$ express the number of degrees of freedom
for the blazar and star, respectively, calculated as the number 
of measurements, $N$, minus 1 ($\nu = N - 1$), while $\alpha$ is the significance level set as
0.1 and 1 percent (i.e $3 \sigma$ and $2.6 \sigma$) in this study.
If the mean $F$ value is 
larger than the critical value ($F_c$), the null hypothesis (i.e., that of no variability) is rejected.

\subsection{$\chi^{2}$-test}

To investigate the presence or absence of variability in our target we also implemented the $\chi^{2}$-test which is interpreted as:

\begin{equation}
\chi^2 = \sum_{i=1}^N \frac{(V_i - \overline{V})^2}{\sigma_i^2},
\end{equation}
where, $\overline{V}$ is the mean magnitude, and the $i$th observation yields a magnitude $V_i$
with a corresponding standard error $\sigma_i$. This error can be attributed to photon noise from the source and sky, CCD read-out and other non-systematic causes.
Calculating exact values of such errors by the IRAF data analysis package is unattainable.
Theoretical errors have been found to be smaller than the real
errors by a factor of 1.3-1.75 (e.g., Gopal-Krishna et al.\ 2003) which for our data is around 1.6, on average. 
Thus the errors obtained after data analysis 
should be multiplied by the above factor to get better estimation of the real photometric errors.
This statistic is then compared with a critical value $\chi_{\alpha,\nu}^2$ where $\alpha$ is the significance level similar to that of
the $F$-test and $\nu = N -1$ is the degree of freedom. $\chi^2 > \chi_{\alpha,\nu}^2$ implies the presence of variability.

%\noindent

\subsection{Percentage amplitude variation}

To characterize the variability of the source in all LCs we calculated the variability amplitude parameter $A$
introduced by Heidt \& Wagner (1996), and defined as
\begin{eqnarray}
A = 100\times \sqrt{{(A_\mathrm{max}-A_\mathrm{min}})^2 - 2\sigma^2}(\%) ,
\end{eqnarray}
where $A_\mathrm{max}$ and $A_\mathrm{min}$ are the maximum and minimum values in the differential LCs of the blazar, and $\sigma$
is the average measurement error.

\subsection{Structure Function}

The Structure function (SF) provides information about the statistical nature of
time series. It is especially well adapted to quantitatively
calculate periodicity and timescales that contribute to
fluctuations, thus providing information on the underlying cause of variability.
The SF has been introduced and discussed at length by
Simonetti, Cordes \& Heeschen (1985). It is not affected by any data gaps in
the LCs and can be applied to unevenly sampled data.

First order SF for a data series is defined as:

\begin{eqnarray}
SF(\tau_j) = \frac{1}{N(\tau)} \sum_{i=1}^N \omega(i) \omega(i+\tau)[a(i)-a(i+\tau_j)]^{2}
\end{eqnarray}
where $\tau$ is the time lag. The weighting function $\omega(i)$ is 1 if we
have observations
for the $i^\mathrm{th}$ interval, else it is 0. Further details on the SF can be found in Gaur et al.\ (2010) and Agarwal et al.\ (2015).
For a sinusoidal time series with period $P$, the SF curve has
minima at $\tau$ equal to the period ($P$) and its sub-harmonics (e.g. Lachowicz et al. 2006).

\subsection{Discrete Correlation Function}

\begin{figure}
\centering
\includegraphics[width=3.5in,height=4in]{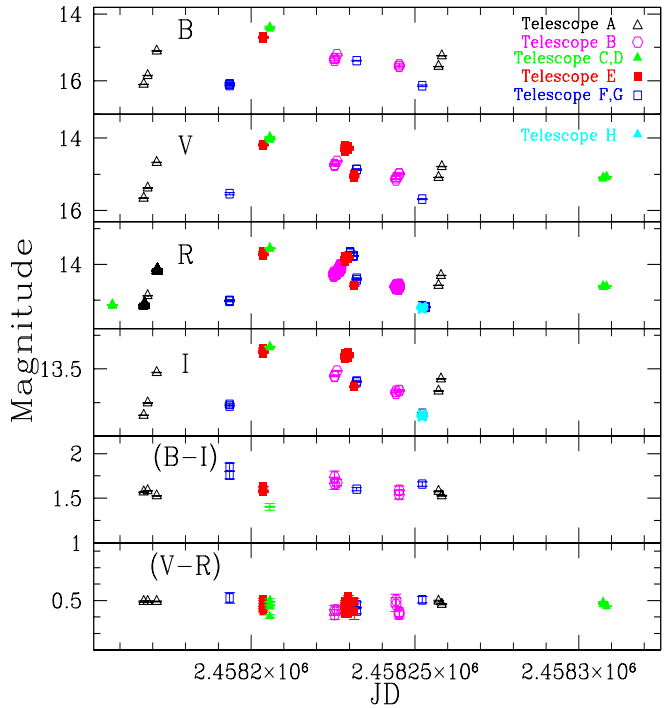}
\caption{Short/Long-term variability LCs and colour indices of 3C\,279 in the $B$, $V$, $R$ and $I$ bands 
and ($B-I$) and ($V-R$) colours. Different colours denote data from different observatories: Black, HCT (Telescope A); Magenta,  ARIES (Telescopes B); Green, JS (Telescope C,D);
Red, HSH (Telescope E);  Blue = Bulgaria (Telescope F,G); Cyan = Turkey (Telescope H). }
\label{stvfigure2}
\end{figure}

To quantify the presence of a periodic signal in the LC of 3C\,279.
if any, we first used the DCF technique
proposed by Edelson \& Krolik (1988). It permits to study a
correlative relationship between any two data sets. For two discrete data
sets ($a_i$, $b_j$ ), we first calculated the unbinned DCF (UDCF) as:
\begin{eqnarray}
 UDCF_{ij}(\tau) = \frac{(a_i-\bar{a})(b_j-\bar{b})}{\sqrt{(\sigma_{a^2} - e_{a^2})(\sigma_{b^2} - e_{b^2})}}
\end{eqnarray}
where $\bar{a}, \bar{b}$ are the mean values of two data sets, \( \sigma_a\), \( \sigma_b \) are their standard
deviations and $e_a$, $e_b$ are measurement errors of data
points in the two data series. Each value of UDCF is associated with
a time delay \( \Delta t_{ij} = (t_{bj}-t_{ai}) \).
The DCF is obtained by averaging the UDCF values for each time lag $\tau$ over the
interval \( \tau - \frac{\Delta\tau}{2} \leq t_{ij} \leq \tau + \frac{\Delta\tau}{2} \)
as following:
\begin{equation}
 DCF{(\tau)}= \frac{\sum_{k=1}^m UDCF_{k}}{M} ,
\end{equation}
where $M$ is the number of pairs with time lag values lying in the $\tau$ interval.
Errors in DCF are calculated using the formula:

\begin{equation}
 \sigma_{DCF(\tau)} = \frac{\sqrt{\sum_{k=1}^{M} (UDCF_k-DCF(\tau))^2}}{M-1} .
\end{equation}

As the two series that were correlated were identical, we obtained the
discrete auto-correlation function (DACF) which was then used to search for
periodicity. The essence of DACF is that for clear correlation, the DACF
peaks at time lags equal to zero, and the presence of periodicity
in the LC will appear as secondary peaks in DACF.

\section{Results}

%To understand the changes in the jet or AD structure and their interaction with the surrounding environment, variability
%studies on timescales of hours to months are of great importance.

Observations of the blazar were carried out for 24 nights between 2018 Feb and 2018 July. Observation log is given in Table 2.
To investigate intraday variability properties, we observed the blazar for $\sim$ 3 -- 5 hours in R band
on a total of 9 nights. Calibrated intraday LCs for our source are shown in Figure 1. In order to statistically examine R band intraday LCs for presence or absence of variations, we performed C test, F test
and $\chi^{2}$-tests as discussed in Sections 3.1, 3.2, and 3.3, respectively.
The LC of the blazar is considered as variable (Var) when variability conditions for all the tests are met at the 0.999 level and is said to be non-variable (NV)
if none of these conditions are met. The source was found to be active during the entire monitoring period. 
Owing to small field of view, out of 9 nights, we were not able to perform variability detection
tests on 2018 Feb 08 and 2018 April 18 LCs as we had insufficient standard stars in the field.
Following above criteria, we found the source to be variable on 5 nights i.e. 2018 Feb 17, 21, April 17, May 05 and 06.
To calibrate the blazar LC of April 18,
we found star 1\footnote{https://www.lsw.uni-heidelberg.de/projects/extragalactic/charts/1253-055.html} appropriate to be used as standard comparison star.
Variability amplitude for April 18 LC was found to be 17.90\% and the LC for the same is displayed in Figure 1. Variability amplitudes for our five IDV nights during which
the source was found to be variable ranged from 5.20\% to 13.90\%.
IDV LCs of 3C 279 in Figure 1 illustrate that the source displayed rise and fall in flux levels on many instances during 3-4 hours of continuous monitoring, thus indicating towards the presence of
characteristic timescale of variations. IDV results and variability amplitudes are listed in Table 3 where column 1 gives observation date, column 2 gives the filter in which observations were carried
out, number of data points in a particular filter are given in column 3, results of $C$-, $F$- and $\chi^{2}$-tests are given in columns 4, 5 and 6, respectively, column 7 tells if the source is variable
or not and column 8 gives variability amplitude.

\hspace*{-0.5in}
\begin{table*}
\caption{Results of IDV observations of 3C\,279.} 
\textwidth=7.0in
\textheight=10.0in
\vspace*{0.2in}
\noindent
\begin{tabular}{cccccccc} \hline \nonumber

 Date       & Band   &$N$  & $C$-test   & $F$-test &$\chi^{2}$test  &   Variable    &$A\%$ \\
 (yyyy mm dd) &        &      & $C_{1},C_{2},C$     &$F_{1},F_{2},F,F_{c}(0.99),F_{c}(0.999)$ &$\chi^{2}_{1},\chi^{2}_{2},\chi^{2}_{av}, \chi^{2}_{0.99}, \chi^{2}_{0.999}$  & & \\\hline 

 2018.02.17   & $R$     & 45  & 5.95, 6.20, 6.08  & 35.41, 38.47, 36.94, 2.04, 2.60  & 1541.5, 1268.2, 1404.8 , 21.7, 27.9  & Var & 6.10 \\
 2018.02.21   & $R$     & 57  & 6.17, 5.83, 6.00  & 38.11, 34.02, 36.07, 1.88, 2.32  & 1496.0, 3079.2, 2287.6, 83.5, 94.5   & Var & 5.20 \\
 2018.04.16   & $R$     & 47  & 1.60, 1.11, 1.36  & 2.56, 1.24, 1.90, 2.01, 2.54     & 90.2, 176.9, 133.5, 71.2, 81.4       & NV &  -- \\
 2018.04.17   & $R$     & 52  & 2.54, 2.60, 2.57  & 6.46, 6.70, 6.58, 1.94, 2.42     & 267.0, 1091.8, 679.4, 77.39, 87.97   & Var & 9.10 \\
 2018.05.05   & $R$     & 84  & 1.80, 2.27, 2.03  & 3.25, 5.14, 4.20, 1.67, 1.99     & 215.5, 1088.8, 652.1, 115.9, 128.6   & Var & 9.10 \\
 2018.05.06   & $R$     & 78  & 3.92, 3.17, 3.55  & 15.43, 10.05, 12.74, 1.71, 2.04  & 865.2, 1816.7, 1340.9, 108.8, 121.1  & Var & 13.90 \\
 2018.05.13   & $R$     & 23  & 1.88, 2.04, 1.96  & 3.53, 4.17, 3.85, 2.78, 3.98     & 1.8, 2.2, 2.0, 40.29, 48.27          & NV & -- \\
              & $I$     & 17  & 2.43, 2.16, 2.30  & 5.90, 4.68, 5.29, 3.37, 5.20     & 2.8, 2.3, 2.5, 32.0, 39.2            & NV & -- \\            
\hline
\end{tabular} \\
\noindent
Var : Variable, NV : Non-Variable     \\
\end{table*}

We found noticeable short term variability (STV) in case of $B$, $V$, $R$, and $I$ passbands.
Our source seems to have reached the faintest state in $B$, $V$, $R$, and $I$ filters on 2018 May 13 as clearly evident from the LC in Figure~\ref{stvfigure2}.
The flux from the blazar was found to increase after 2018 May 13 for
the next few days.
During our observation run, 3C\,279 reached the brightest $R$ band magnitude
of 13.54 which is just $\sim 0.94$ mag fainter than its flux level of $R \sim 12.6$ reported by Gupta et al.\ (2008) when the source
was in an outburst state. The source decayed significantly reaching $R \sim 15.25$, which is still brighter than its faint state of $R \sim 17.1$ reported by Rani et al.\ (2010).
Above results are summarized in Table 4.

To calculate the variability amplitude in each filter during the entire monitoring period, we used Equation 4. 
The short term variability amplitude was found to increase with frequency, with the following values: $\sim 177$,
  $\sim 172$, $\sim 171$, and $\sim 158$ per cent in $B$, $V$, $R$, and $I$ bands, respectively
which is in accordance with other investigators (Ghisellini et al.\ 1997;
Papadakis et al.\ 2003; Bonning et al.\ 2012). Such a trend is dominant when substantial variability is present over the observation duration.
A few possible scenarios of optical emission include the standard shock-in-jet model (e.g., Marscher \& Gear 1985; Spada et al.\ 2001; Joshi \& Bottcher 2011),
colliding plasma shells (e.g., Guetta et al.\ 2004), variations in the direction of forward beaming (e.g. Villata \& Raiteri 1999) and many more.
The most promising scenarios for flux variations in blazars from intraday to long timescales is the shock-in-jet model where shocks from the base of jet traveling down
the Doppler boosted relativistic jet that induce significant flux fluctuations,
accelerating particles and/or compressing magnetic field (Marscher 2014).
Thermal emissions in the optical scenario could be associated with the AD instabilities such as hotspots when the source is in the low brightness state (Mangalam \& Wiita 1993; Chakrabarti \& Wiita 1993).
STV of blazars is also well modelled in terms of the helical jet model (Marscher \& Travis 1996).
Variations in viewing angle might also account for changes in the source brightness. At larger viewing angles,
the source is fainter while at a smaller angle it is brighter (Lainela et al.\ 1999).
We also investigated corresponding ($B-I$) and ($V-R$) variations on short term basis. Colour variations with time are displayed in the lower two panels of Figure~\ref{stvfigure2}.
The variability amplitude for ($B-I$) was calculated to be 0.47\% with
a maximum value of 1.84 and a minimum of 1.38~mag, while the amplitude of variability for ($V-R$) was estimated to be 0.12\% for a maximum
of 0.52 and a minimum of 0.40~mag. Larger ($B-I$) values are
expected owing to increase in standard deviation with frequency separation between two bands.

\begin{figure*}
\epsfig{figure=  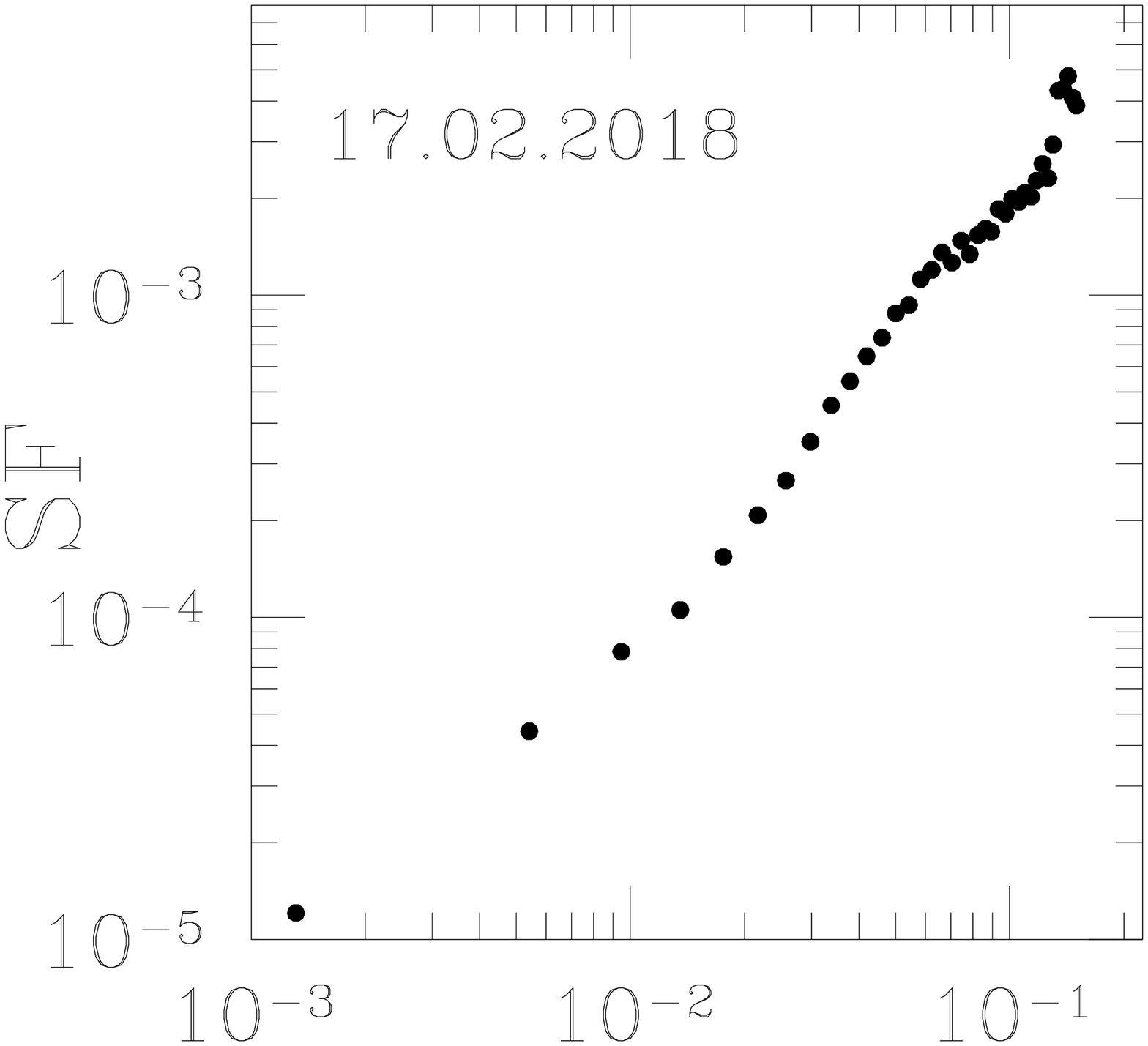,height=2.1in,width=2in,angle=0}
\epsfig{figure=  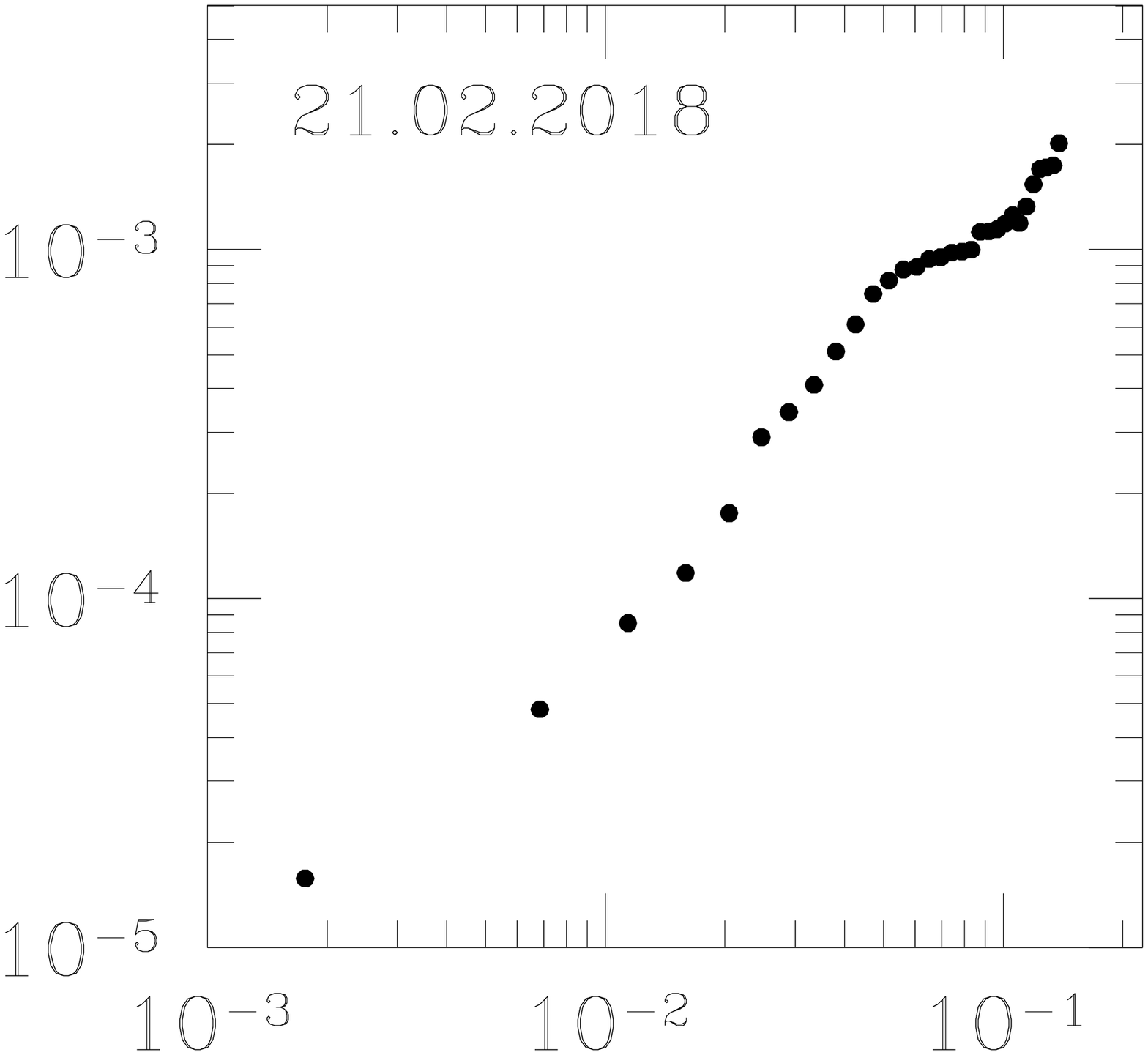,height=2.1in,width=2in,angle=0}
\epsfig{figure=  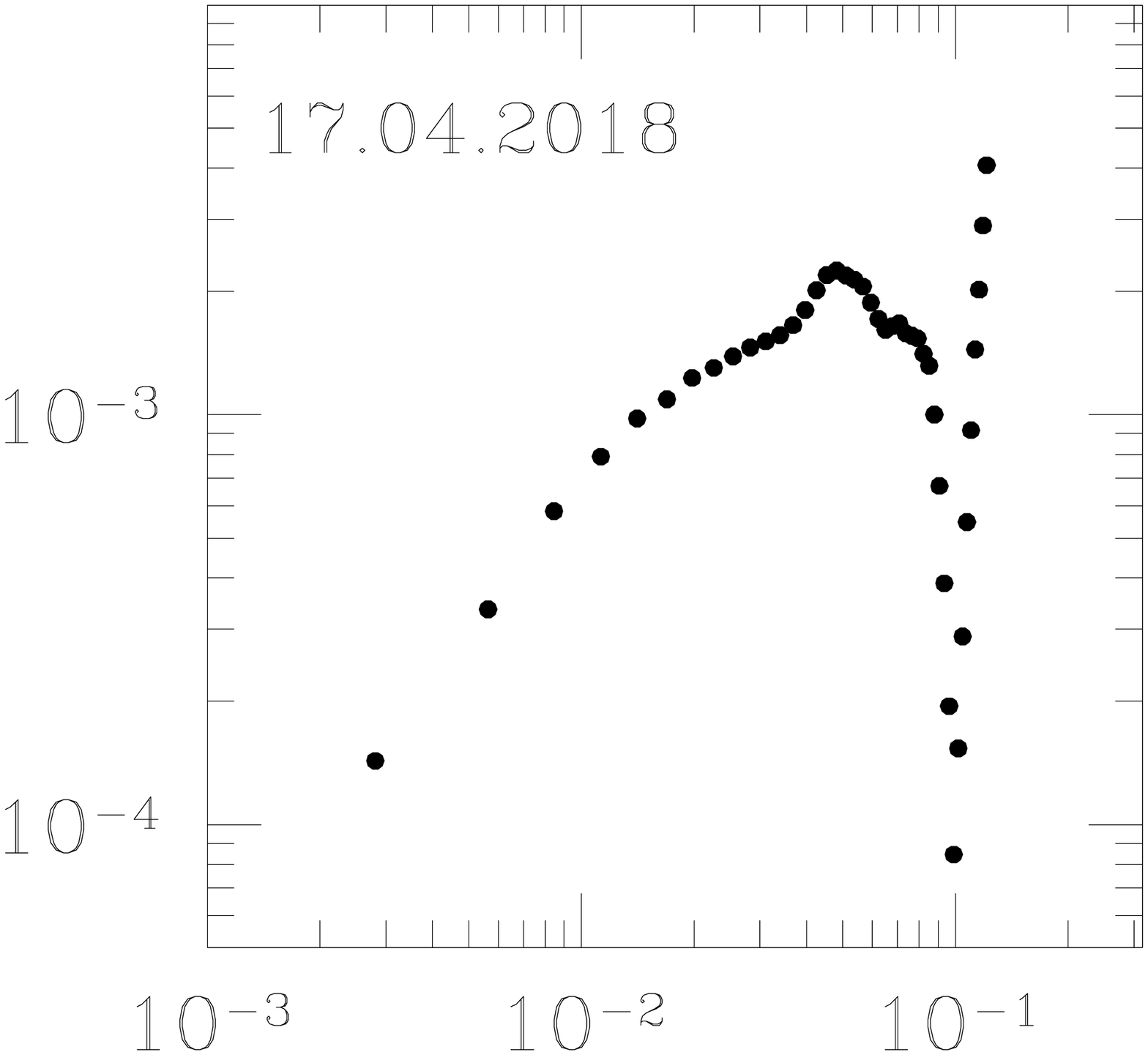,height=2.1in,width=2in,angle=0}
\epsfig{figure=  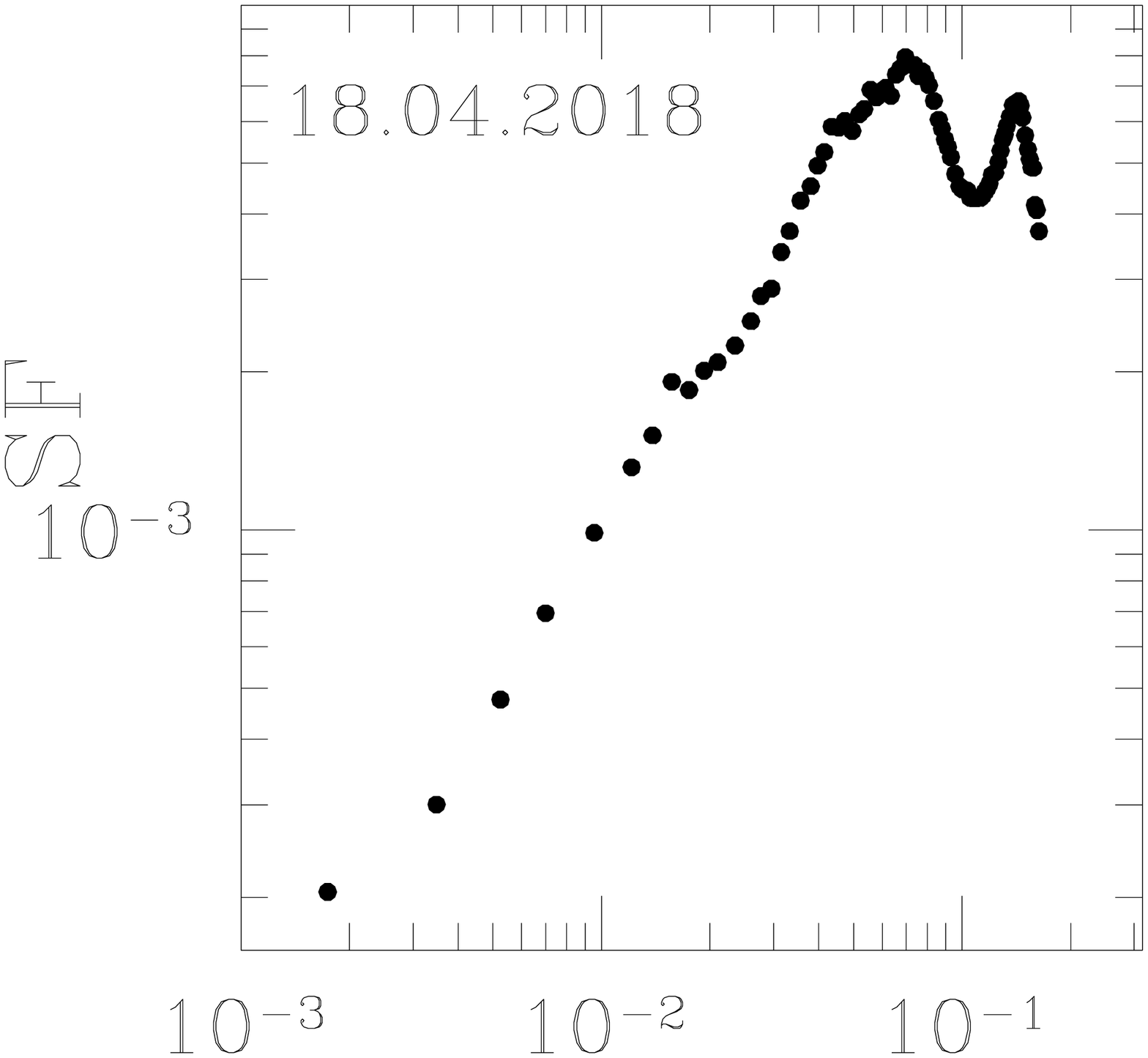,height=2.1in,width=2in,angle=0}
\epsfig{figure=  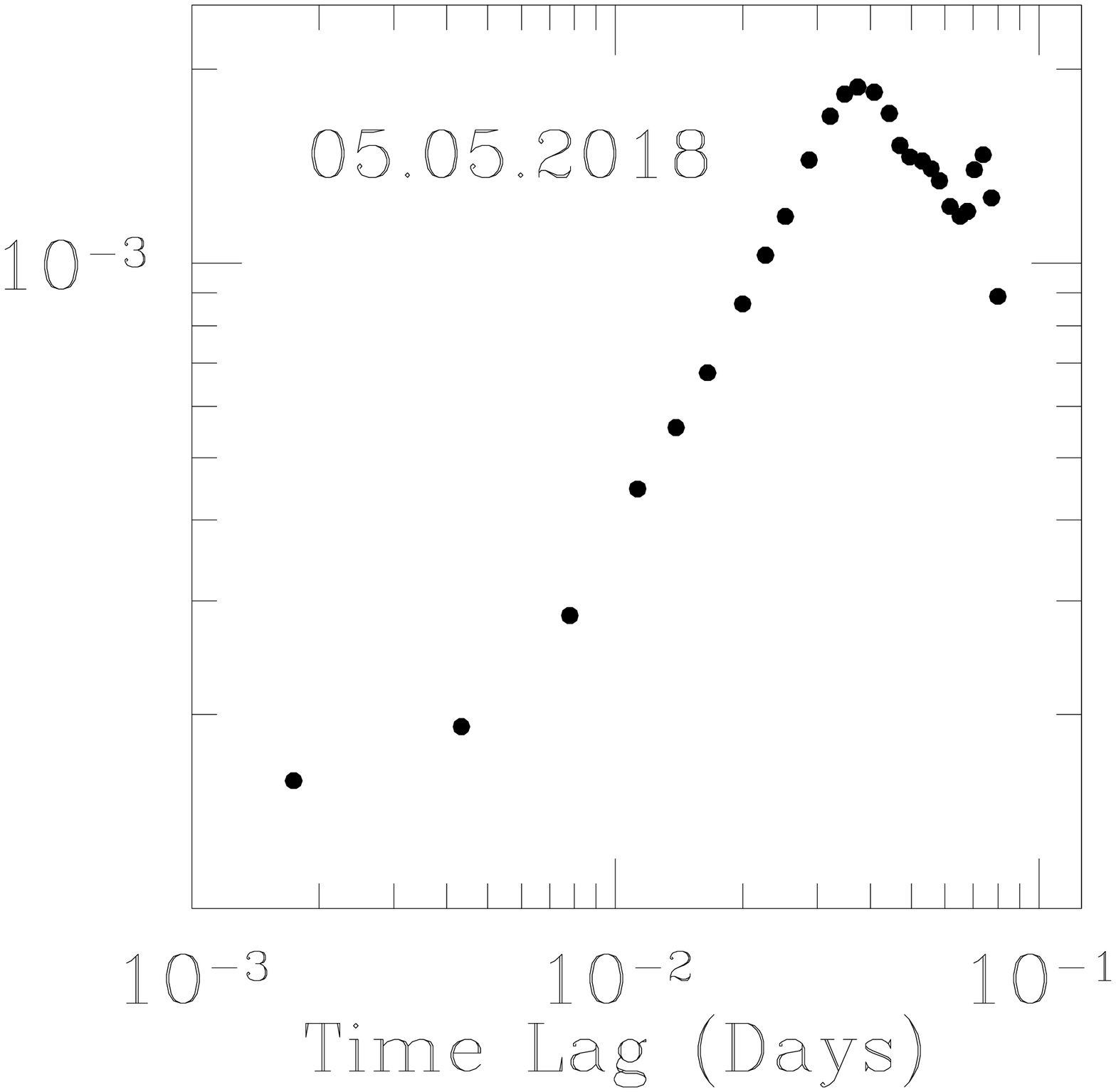,height=2.1in,width=2in,angle=0}
\epsfig{figure=  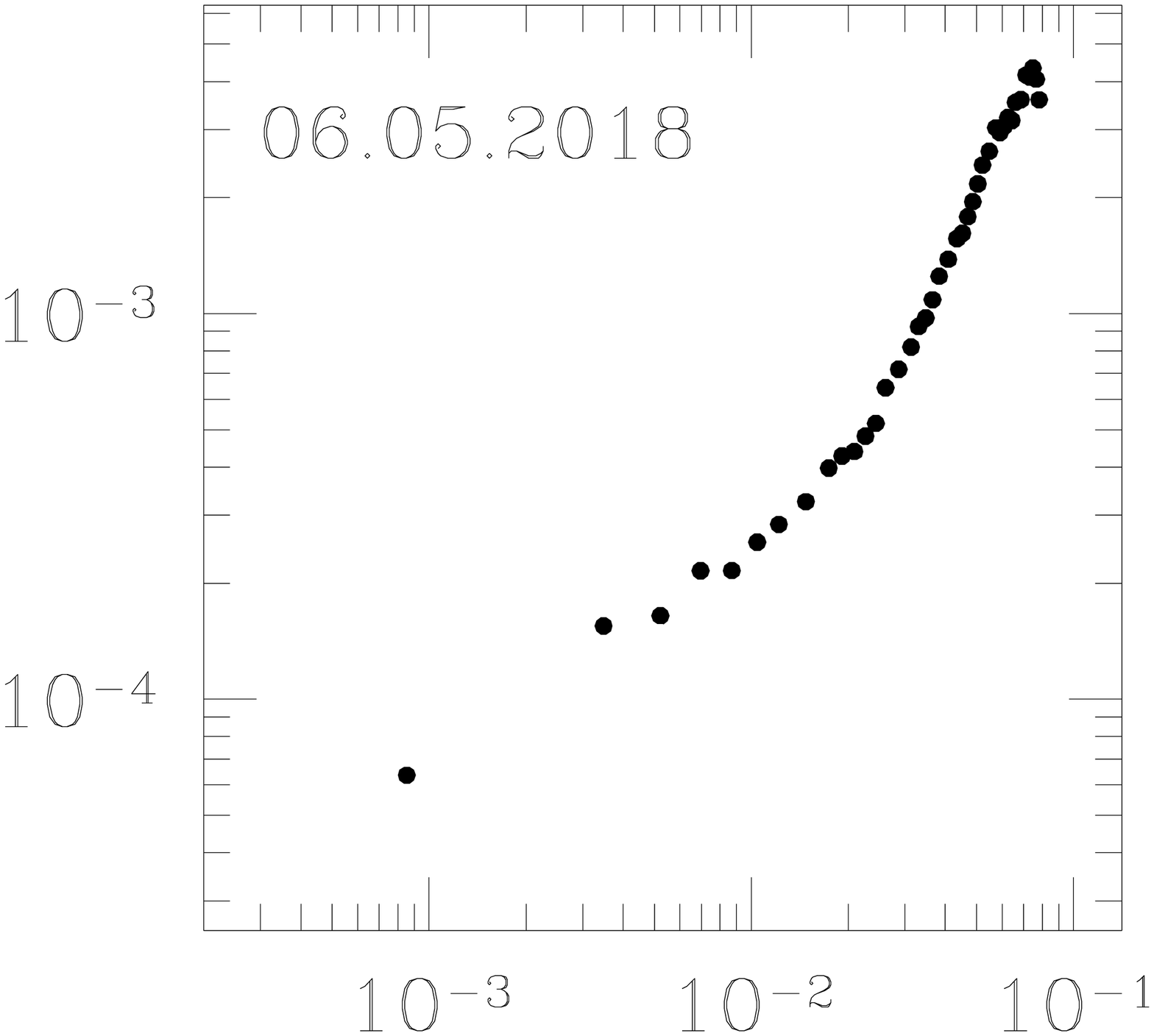,height=2.1in,width=2in,angle=0}
  \caption{SF plots for the blazar 3C\,279 in the $R$
    passband. Observation date is indicated in each plot. In each plot, X and Y axis are the time lag (days) and SF values, respectively.}
\label{SF_BL}
\end{figure*}

\begin{figure*}
\epsfig{figure=  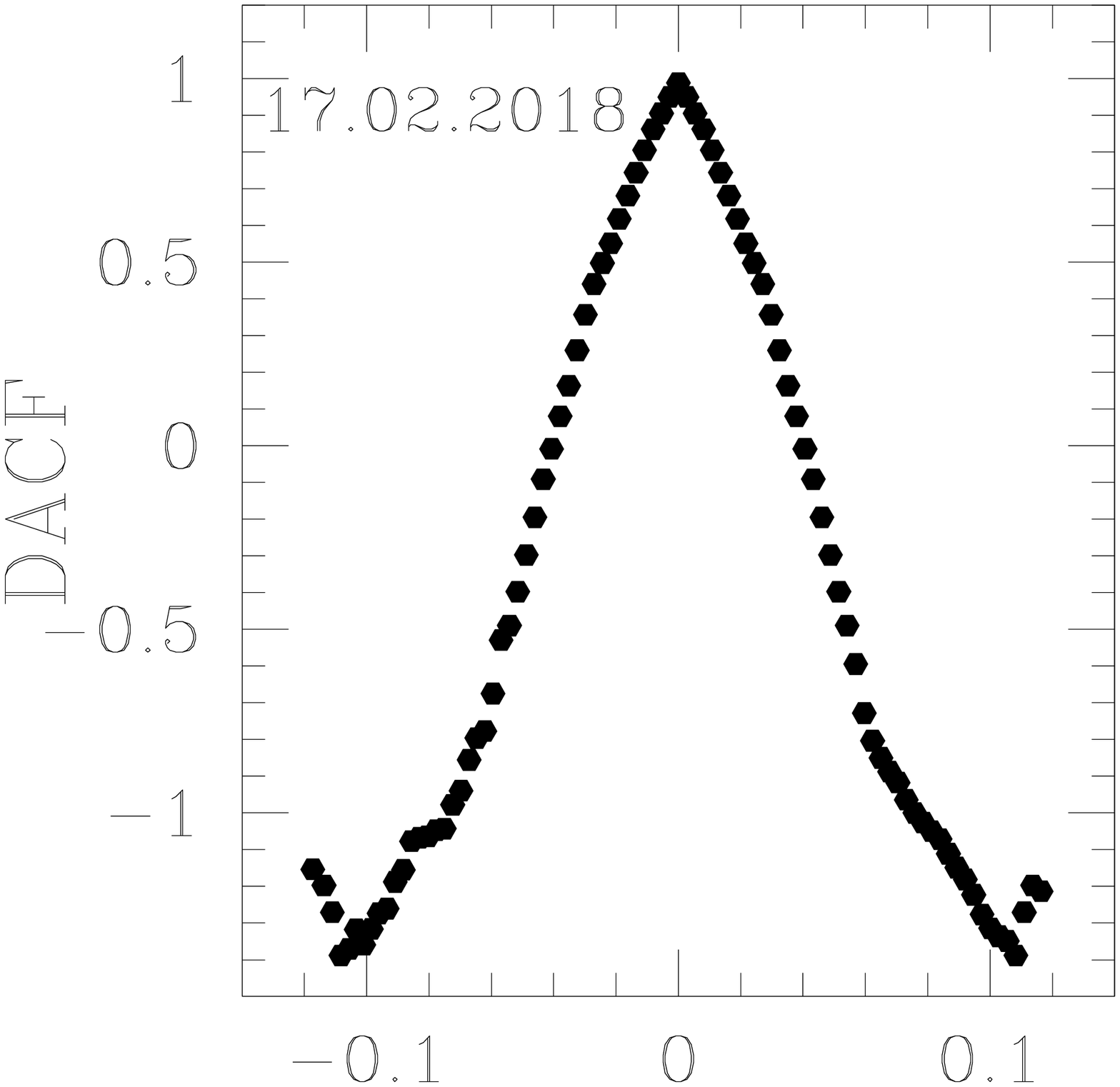,height=2.1in,width=2in,angle=0}
\epsfig{figure=  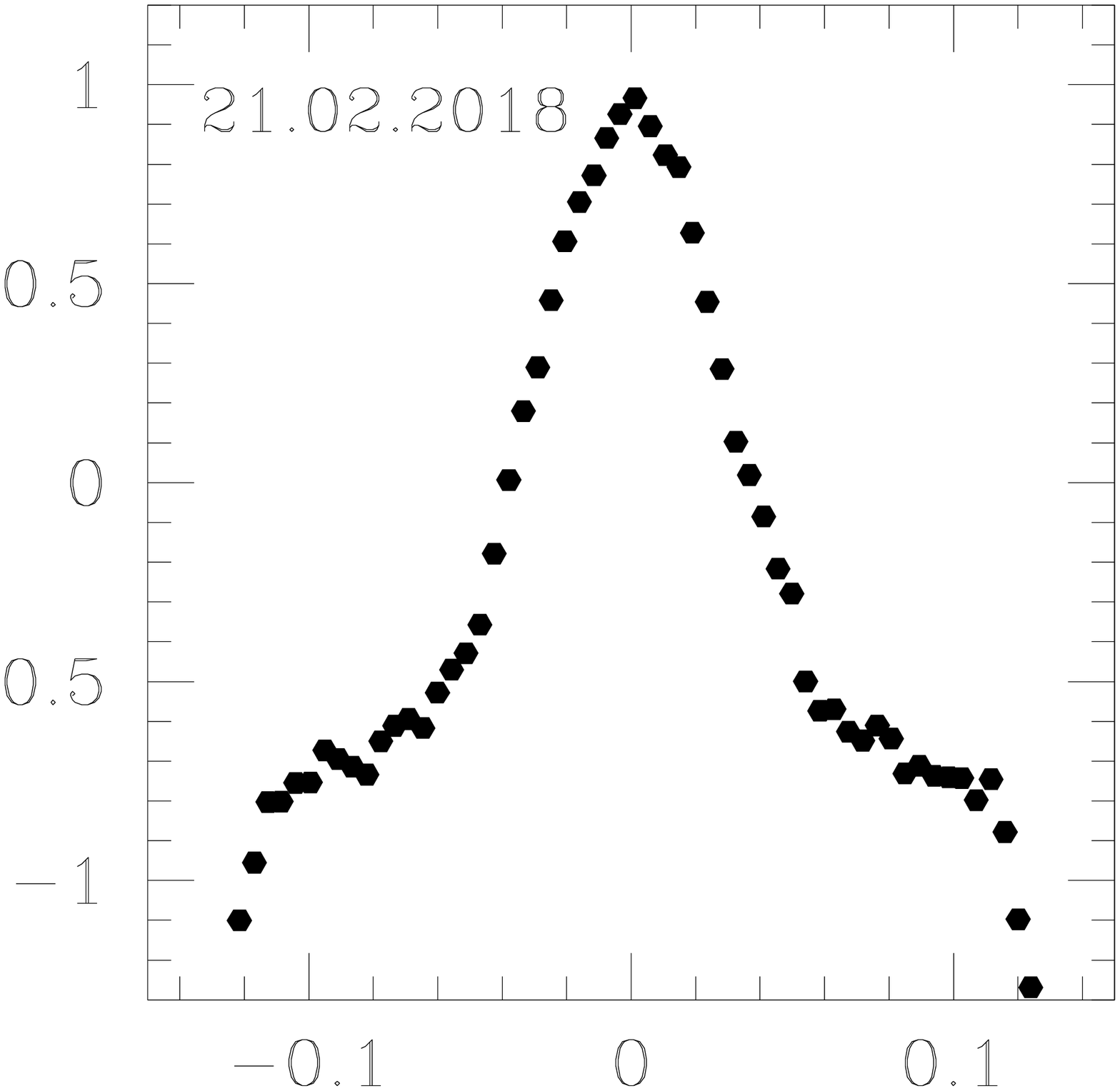,height=2.1in,width=2in,angle=0}
\epsfig{figure=  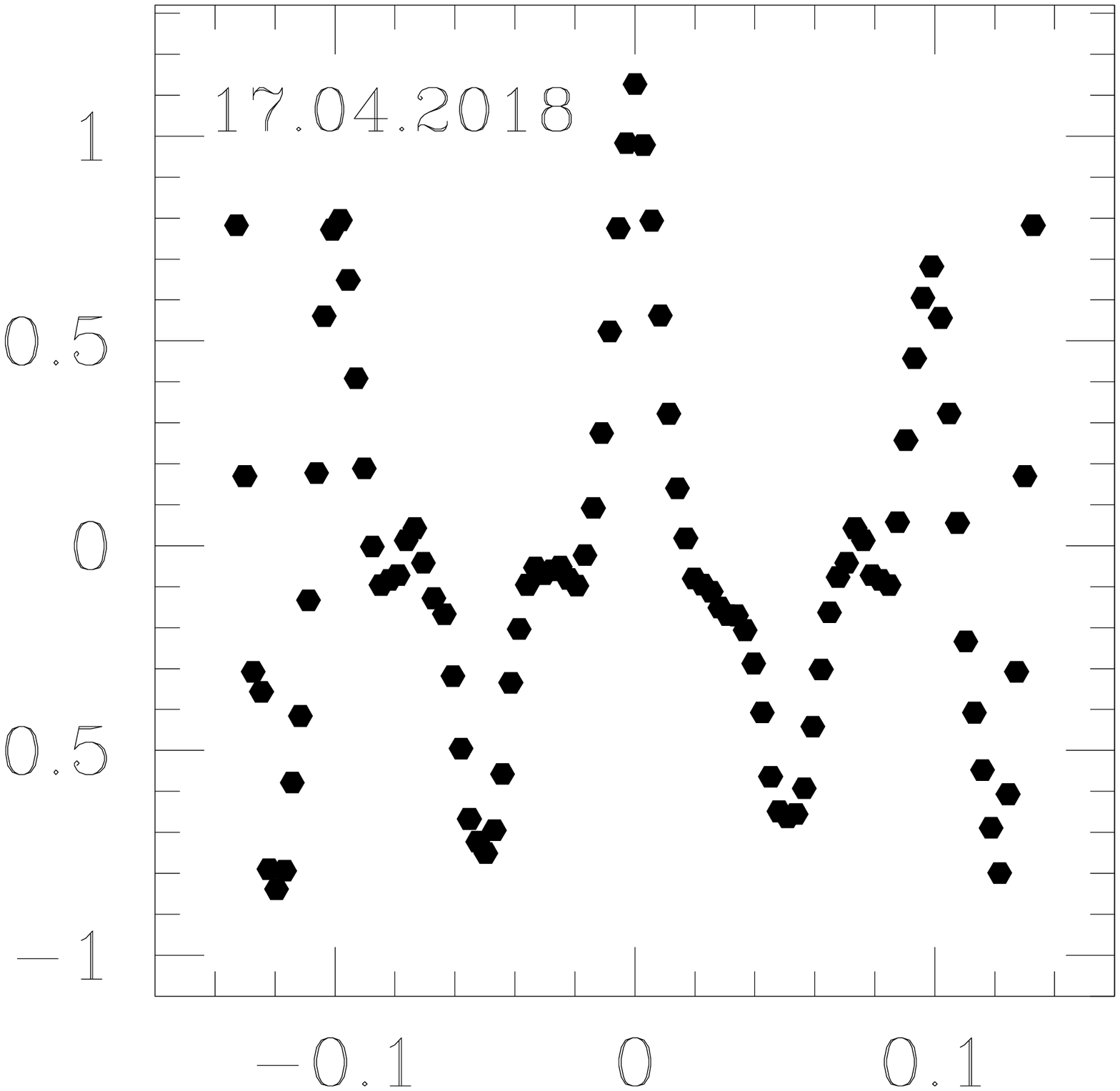,height=2.1in,width=2in,angle=0}
\epsfig{figure=  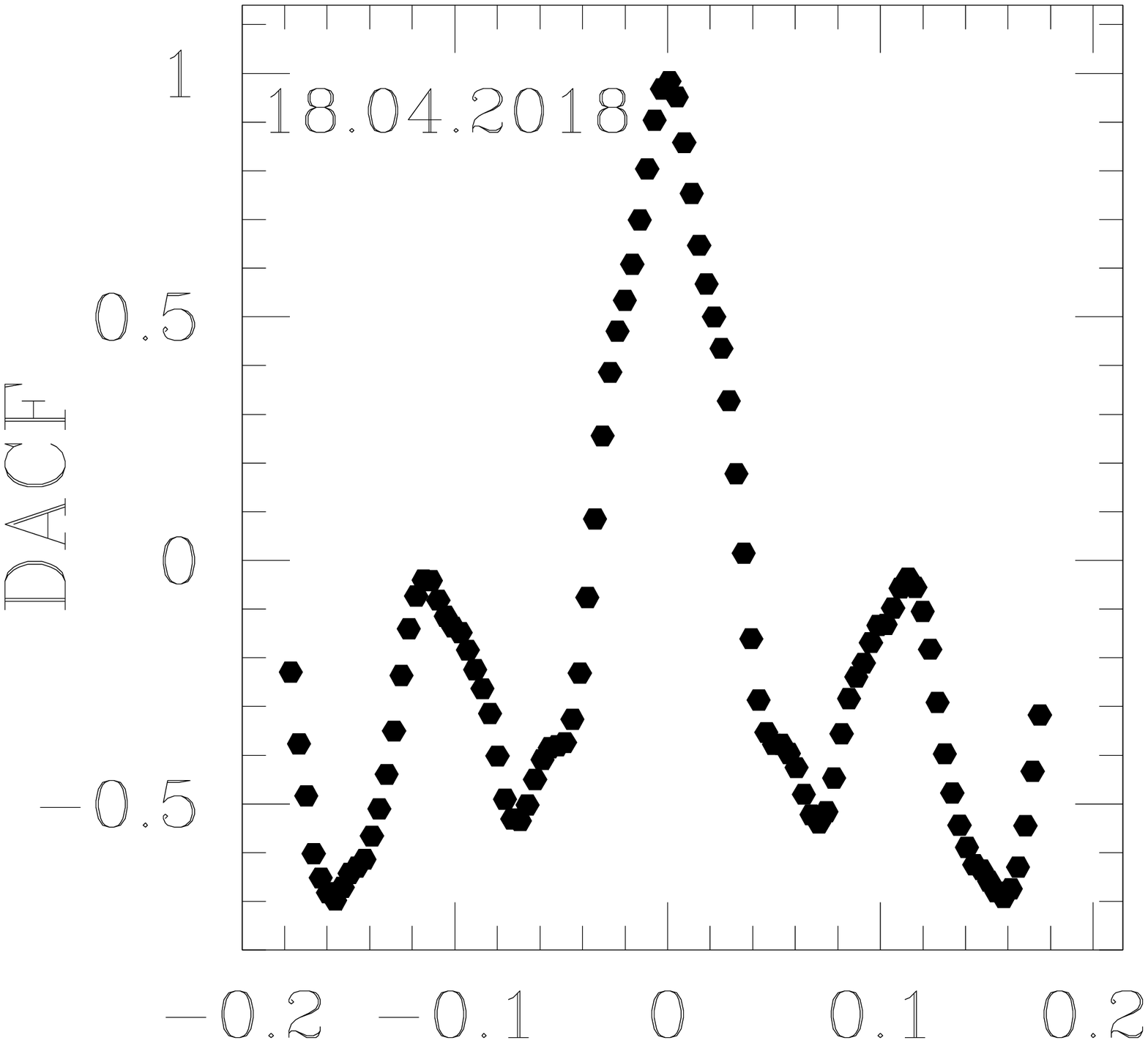,height=2.1in,width=2in,angle=0}
\epsfig{figure=  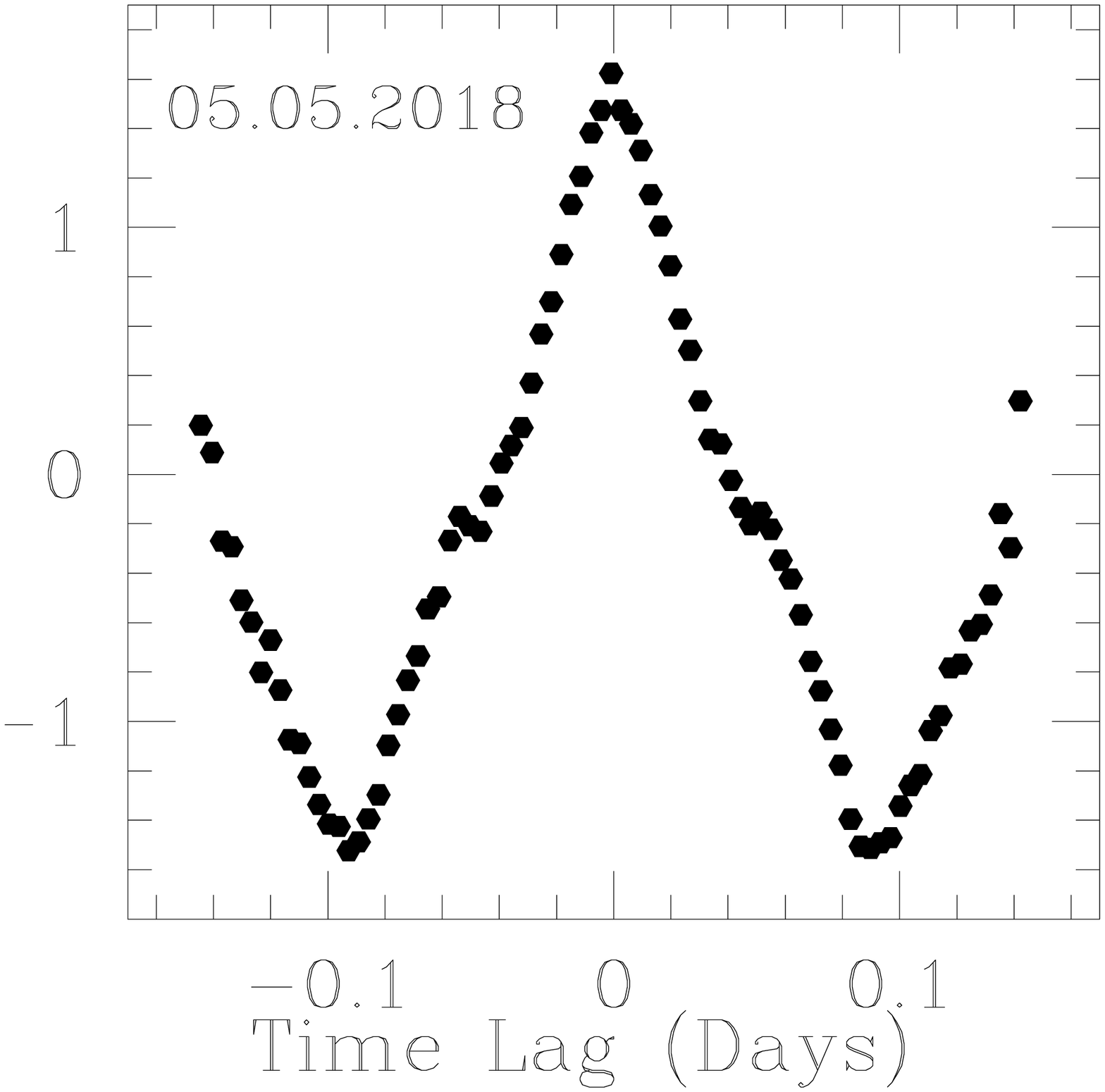,height=2.1in,width=2in,angle=0}
\epsfig{figure=  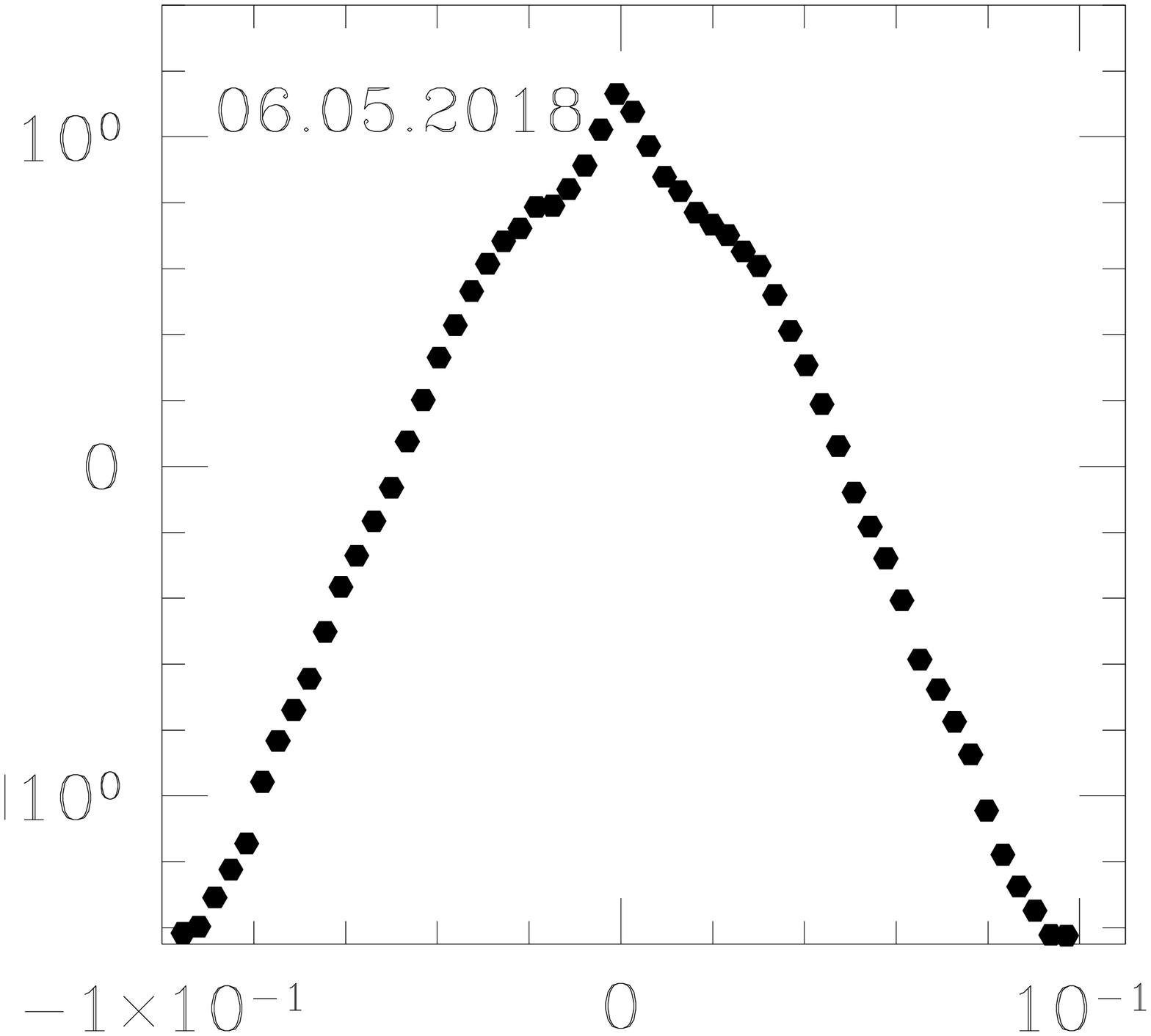,height=2.1in,width=2in,angle=0}
  \caption{DACF plots for the blazar 3C 279 in R passband. Observation date is indicated in each plot.
In each plot, X and Y axis are the time lag (days) and DACF values, respectively.}
\label{ACF_BL}
\end{figure*}

\begin{table}
\caption{Results for STV studies displaying magnitude changes in each band.  }
\textwidth=6.0in
\textheight=9.0in
\vspace*{0.2in}
\noindent
\begin{tabular}{p{0.25cm}p{0.5cm}p{1.3cm}p{0.6cm}p{1.1cm}p{2cm}} \hline
 Band & Faintest  & ~~~Date  & Brightest & ~~~Date  \\ 
      & ~~Mag       & ~~~(Max) &~~Mag  & ~~~(Min)  \\ \hline 
 $B$    & 16.15  & 2018.05.13 & 14.38 & 2018.03.28  \\
 $V$    & 15.68  & 2018.05.13 & 13.96 & 2018.03.28  \\
 $R$    & 15.25  & 2018.05.13 & 13.54 & 2018.03.28  \\
 $I$    & 14.58  & 2018.05.13 & 12.99 & 2018.03.28  \\
 \hline
\end{tabular}  \\
Column 1 indicates the band in which observations were taken,
column 2 represents the maximum magnitude attained by the source in the filter given in column 1 on a particular date, which is mentioned
in column 3, followed by the minimum magnitude value and respective date in columns 4 and 5.
\end{table}

\subsection{Variability timescales}

To quantitatively calculate timescale of variations in the optical fluxes we used SF and DACF techniques as explained in Section 3.5 and 3.6. We have constructed DACFs and SFs for all those observation dates
when the source was found to be variable. SFs are displayed in Figure~\ref{SF_BL} while DACFs are shown in Figure~\ref{ACF_BL}.

The SF for 2018 Feb 17 displays a monotonic increase with no detectable
plateau which implies that variability timescales are longer than the observation span. Similarly, the DACF also did not display any
significant trend.
For Feb 21, we have a double hump appearance in the SF plot at $\sim$ 87 minutes and $\sim$ 202 minutes. As evident from the plot, plateau was not followed by any dips thus
these timescales cannot be considered significant and could be due to photometric and systematic errors on the data points. Similar trend with same timescale values was found by DACF analysis also.
The LC for the night of April 17, shows indications of 3 humps and 2 dips with characteristic variability timescales of 72, 108, 173 minutes. Dips provide evidence of quasi-periodicity in the LC.
To cross check for the presence of these detected timescales of variability we performed DACF analysis. As evident from the Figure 4, similar variability timescales were detected
from DACF analysis along with hints of
quasi-periodicity. The nominal timescales of variability for April 18 were found to be about 23, 100 and 200
minutes.
Similar trends were suggested by SF plot of May 05 displaying 2 plateaus and dips.
For May 05, we detected possible variability timescales of $\sim$ 60 and 120 minutes.
The first dip of the SF might hint towards periodicity in the LC, but as the subsequent dips are
absent, the detected periodicities cannot be considered significant. The nominal variability timescales for April 18 and May 05 were supported by DACF technique also.
SF plot for May 06 displays a continuously rising trend giving a possible variability timescale of
115 minutes. Since the plateau was not followed by any dips, any variability timescale is greater than or equal to the observation duration. Also, SF results were not supported by DACF analysis and hence
are not reliable.

Variability timescales can be used to find the size of the
emitting region or the Eddington luminosity.
To relate observed quantities with the rest frame quantity we make use of Doppler
boosting factor
\begin{equation}
 \delta = \frac{1}{\Gamma (1-\beta~{\rm cos}\theta)}
\label{eq.delta}
\end{equation}
where $\theta$ is the angle which the LOS makes with the jet axis, $\beta = v/c$, with $v$ being the velocity of the plasma in the jet and $c$ is the velocity of light in vacuum,
while $\Gamma$ is the bulk Lorentz factor of flow which is given as, $\Gamma = [1 - \beta^{2}]^{-1/2}$.
Shortest timescale of variability
is proposed to be associated with the light crossing time. The size of the
emitting region is given as $R \leq c\, \delta \Delta t/(1+z)$.
Chen (2018) obtained Doppler factor of 27.7 by fitting the NASA/
IPAC Extragalactic Database (NED)\footnote{https://ned.ipac.caltech.edu/} generated SED with a one -- zone synchrotron + IC model. 
However, given the non-simultaneity of the SEDs fitted, the estimated parameters of a single object have to be considered with some caution.
Jorstad et al.\ (2017) calculated individual Doppler factors for a set of knots using their kinematic data related to the knots ejected before 2013. The weight-averaging of the individual 
measurements results in a Doppler factor of 15.3 $\pm$ 3.9 (weighted standard deviation of 7.5).
Liodakis et al. (2018) calculated Doppler factor of 11.64$^{+1.11}_{-1.77}$ using a Bayesian approach to model the 2008 -- 2017 
radio curve of 3C 279. This result could be considered as an average Doppler factor for that period and it is in good agreement 
with the mean Doppler factor estimated from the kinematic data. Moreover, the modeled curves are very close in time to our monitoring 
campaign. So, we shall use the so obtained Doppler factor in our further consideration.

For each night we derived the minimal time scale as the first plateau/maximum of SF.
Using the estimated time scales and the Doppler factor of 11.64 we
calculated the upper limits of the size of the regions responsible for the intra-night
variability (Table\,\ref{sizes}). Timescales of variability detected during our observation run ranged from 23 minutes to 115 minutes.
Ackermann et al.\ (2016) detected a significant flux variability at sub-orbital time-scales of $\sim$ 5 minutes using the Fermi-LAT observations. Variability timescales as short as $\sim$ 5 minutes in
gamma-rays can be explained by mirror driven clumpy jet model or/and model based on synchrotron origin from a magnetically dominated jet.
More detections of minute scale variability timescales will help to disentangle the theory behind various blazar emission models.

The estimated sizes of the emitting regions for 3C\,279 are consistent with those found for other blazars.
For example, Kaur et al. (2017) found sizes in the range $(7.0 \times 10^{14} - 3.5 \times 10^{15}) \,\rm cm$ over a period of 10 years for 3C\,66A.
The continuous, 72\,hours long LC of S5\,0716+714 (Bhatta et al. 2013) provided an unique opportunity to model flares with synchrotron pulses 
and to estimate sizes of the turbulent cells in a consistent way~-- the sizes reported cover a range from $9.0 \times 10^{13}\,\rm cm$ to $2.5 \times 10^{15}\,\rm cm$. 
Rafle et al. (2012) applied the above model to $\sim$6 years long campaign on S5\,0716+714 (Montagni et al. 2006) and obtained the cell sizes in the
range $(6.0 \times 10^{13} - 1.0 \times 10^{15})\,\rm cm$.
Based on these literature estimates (in addition to ours) we could claim that the typical size of the regions responsible
for the intra-night variability lies in the range $(6.0 \times 10^{13} - 3.5 \times 10^{15})\,\rm cm$. 
The precision of the these limits could be further increased if we enlarge the number of the independent size estimates. 
This, however, is beyond the scope of this paper.
In addition, the modeling of Rafle et al. (2012) and Bhatta et al. (2013) revealed that almost all of the turbulent cells are with sizes less than $\sim$$7.5 \times 10^{14}\,\rm cm$.

The turbulence is a stochastic process and each intra-night LC is a particular realization of this process. 
Therefore, increasing of the number and quality of intra-night LCs we could gain a knowledge about the turbulence
in the relativistic plasma. In the framework of Kolmogorov theory the smallest scales probe the regions where the 
viscous dissipation of turbulence kinetic energy takes place (the so called Kolmogorov scale, a smallest scale in a turbulent flow);
the largest scales mark the regions of the energy injection in the turbulent region. Based on the above considerations we could state that the
upper limit for the Kolmogorov scale in blazar jets is $6.0 \times 10^{13}\,\rm cm$. This is a rough estimate and it should be
made more precise increasing the number independent size estimates.

The sizes of the regions responsible for intra-night variability are much smaller than those used in
the the SED modeling, which are typically of about $10^{17}\,\rm cm$ (e.g., Banerjee et al. 2019).
In any case the maximal intra-night time scales set an lower limit on the jet size.

\begin{table}
\caption{Time scales of the intra-night variability and the corresponding upper limits of the emission region sizes (AU = Astronomical Units).}
\centering
\begin{tabular}{cccc}
\hline
Date of observations & Time scale & \multicolumn{2}{c}{Size} \\
\cline{3-4}
% \multicolumn{4}{}{} \\
\noalign{\smallskip}
        (yyyy mm dd) &      (min) & ($\times 10^{15}\,\rm cm$) & (AU) \\
\hline 
2018 02 21 &  87 & $1.19 \pm 0.18$ & $ 79 \pm 12$ \\
2018 04 17 &  72 & $0.98 \pm 0.15$ & $ 66 \pm 10$ \\
2018 04 18 &  23 & $0.31 \pm 0.05$ & $ 21 \pm  3$ \\
2018 05 05 &  60 & $0.82 \pm 0.12$ & $ 55 \pm  8$ \\
2018 05 06 & 115 & $1.57 \pm 0.24$ & $105 \pm 16$ \\
\hline
\end{tabular}
\label{sizes}
\end{table} 

\subsubsection{The composite April 18 flare}

Among the intra-night light curves shown in Figure 1 the most complicated is that one on April 18.
The other multi-peaked LCs were observed on April 17 and May 05. The April 17 LC shows somewhat
flat-topped flares, which can be a result of the single flares overlapping, whereas the May 05 LC
is too noisy. So, we shall consider in details only the April 18 LC.

At the beginning of the monitoring on April 18 the source fades by $\sim$0.7\,mag within $\sim$30\,min. This fading
is followed by 4 overlapping flares. After the last flare~-- near the end of the monitoring~-- the flux
slightly increases and shows some fluctuations. We, however, cannot make a firm conclusion what
this feature is because of the end of the observing set. The first flare peaks at $\sim$17.4 hours UT and the effect of the flare onto the SF could
be seen as a local maximum at $\sim$0.015 days or $\sim$23 minutes. To get information about the characteristics of the flares
we decomposed the intra-night light curve using a sum of 4 double exponential functions (see Abdo et al. 2010).
We excluded from the fit the initial fading and the final brightening of the source and assumed the
flux level underlying the flares to be constant. The flares were fitted simultaneously using the
weighted least-squares fitter {\tt MPFIT} (Markwardt 2009). The decomposition is shown in Fig.\,\ref{lc:fit}.
In the further discussion we shall exclude the decay time scale of the first flare and the rise time scale of the third flare
owing to the small number of data points covering the corresponding phases of the flares.

The fitted rise ($e$-folding) time scales of the individual flares were found to be consistent to each other to within the formal uncertainties.
The same applies for the decay time scales as well.
In addition, the decay time scales were found to be systematically larger than the rise ones yet consistent with them to within the uncertainties.
We derived the weighted mean (over the individual flares) rise and decay time scales in the observer's frame to be $(13.5 \pm 2.5)$\,min,
$\chi^2_{\rm df}=0.7$, and $(14.6 \pm 6.0)$\,min, $\chi^2_{\rm df}=0.9$, 
respectively, where the uncertainties quoted are the weighted standard deviation about the weighted mean.
We shall assume that the flares are symmetric based on these weighted mean results.
The obtained mean rise time scale is consistent with that one obtained if we consider the local maximum
on the corresponding SF the later being, however, more conservative estimate. The flares symmetry could mean either the injection
time is comparable, while the cooling time is shorter than the light
crossing time (Chiaberge \& Ghisellini 1999) or geometric effects
are in play. Unfortunately, this issue cannot be resolved owing the
lack of multiband light curves for April 18.

Let us assume that the individual flares are produced by turbulent cells which
cool by synchrotron emission after being hit by a strong shock (see below). 
Then the multiple overlapping flares observed on April 18 could mean that the shock hits a fragmented region within the jet, consisting of
at least 4 distinct cells; it is worth mentioning in this context Rafle et al. (2012) who assumed the outliers in their cell sizes distribution could be an unresolved group of cells rather than a single one. 
We obtained an upper limit of the size of this region as $(1.24 \pm 0.10)\times10^{15}\,\rm cm = (84 \pm 6)\,\rm AU$ given the estimated size of a single emitting cell (cf. Table\,\ref{sizes}).
The region size is consistent with the upper limit derived in the previous section.

This kind of multiple fast flares that overlap is not commonly observed during the intra-night monitoring campaigns. One of the most spectacular case was captured by Man et al. (2016):
they detected 5 flares within $\sim$70 minutes. The colour-magnitude diagram for the composite flare showed strong BWB trend with a hysteresis (Zhang et al. 2016):
a typical feature predicted within the shock-in-jet scenario for blazar's intra-night variability (e.g., Kirk et al. 1998). The BWB trend observed for these flares makes our assumption about
the non-geometrical origin of the composite April 18 flare reasonable.
The study of such multiple fast flares is of great importance as they probe the smallest jet scales where the energy dissipates.

\begin{figure}
\centerline{\includegraphics[width=2.5in]{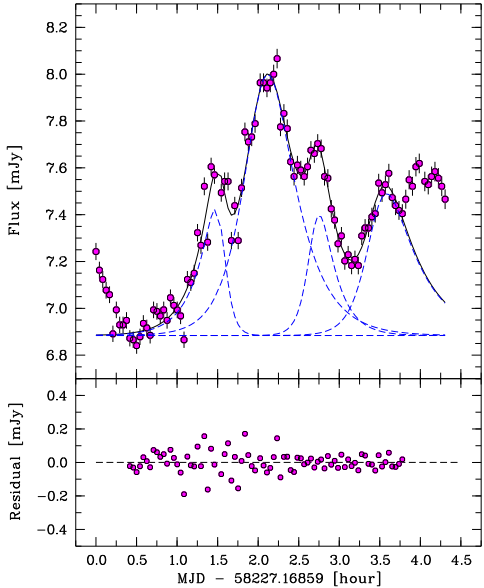}}
\caption{Decomposition of the April 18 flare. The fitting residuals have a standard deviation of 0.06\,mJy.}
\label{lc:fit}
\end{figure}

\subsection{Colour-magnitude relationship}

Optical flux variations are accompanied by spectral changes, thus resolving colour-magnitude (CM) relationship can be useful in understanding the origin of blazar emission and also explore various
variability scenarios. 
We now look for any relationship between the colour indices of the source and the brightness in the $V$ band.
We fitted the plots of colour indices (CIs) vs. $V$ magnitude with straight lines i.e (CI $= m\,V + c$) and calculated the fit values of the slope, $m$,
the constant, $c$, the Spearman correlation coefficient $r$ along with the corresponding null 
hypothesis probability, $p$ which are listed in Table 6. A positive slope
implies significant positive correlation (when the null hypothesis
probability is $p \le 0.05$) between CI and blazar $V$ magnitude,
which in turn implies that the source exhibits a bluer when brighter (BWB) or redder when fainter trend (H.E.S.S.~Collaboration et al.\ 2014), while a negative slope implies redder when brighter (RWB)
trend. A significant positive correlation between $V$ band magnitude and colour indices on few months timescales was detected for ($R-I$) and ($V-R$),
while no clear trend was observed for ($B-I$) and ($B-V$),
as evident from Table 6. A CM plot for short timescales is shown in Figure 6. In Figure 6, offset values of 1.0, 1.3, and 0.5 are used with ($B-I$), ($B-V$) and ($R-I$), respectively, for clarity.

Colour behaviour obtained by us is similar to Raiteri et al.\ (2003). They also found weak BWB trends on few instances only.
BWB trend has been predominantly observed in blazars by most of the optical observations (Ghosh et al.\ 2000; Clements \& Carini 2001; Gu et al.\ 2006; Rani et al.\ 2010; Agarwal \& Gupta 2015;
Gupta et al.\ 2016) which was also exhibited by our quasi-simultaneous observations.
As given by Sasada et al.\ (2010), CM relationship in blazars varies among different states of the source i.e. outburst state, active state, and faint state.
During our observation run, we found that the source attained the maximum flux level of 13.537~mag in $R$ pass band, which is just $\sim 0.94$~mag fainter than its brightest known magnitude of $R \sim 12.6$
(Gupta et al.\ 2008). Therefore, most likely we observed the source in
a post-outburst state. Among the above mentioned three states, we can best describe the source as being in active state and thus the
colour trend obtained by us is in agreement with those reported by Sasada et al.\ (2010).
No clear trends have been detected by several authors (Stalin et al.\ 2006; B{\"o}ttcher et al. 2007; Agarwal et al.\ 2016).

The average spectral indices are calculated using the following relation (e.g., Wierzcholska et al.\ 2015),
\begin{equation}
\langle \alpha_{BR} \rangle = {0.4\, \langle B-R \rangle \over \log(\nu_B / \nu_R)} \, ,
\end{equation}
where $\nu_B$ and $\nu_R$  are effective frequencies of the respective bands (Bessell, Castelli, 
\& Plez 1998).

The optical slope for all nights, when quasi-simultaneous observations in $B$, $V$, $R$, and $I$ filters were taken, were calculated using equation 10 and were found to range between $2.3 \pm 0.05$
to $3.0 \pm 0.03$.
These steep spectral index values imply strong synchrotron emission from the Doppler boosted blazar jet.
Optical emission from blazars can be explained
by different theoretical models, namely: AD based model and the shock-in-jet model.
According to the Blandford \& Rees (1978) model, non-thermal emission in case of blazars is associated with the relativistic Doppler boosted jet pointed towards
the observer's direction.
Therefore, optical emission from blazars can be attributed to the shock-in-jet models. It is expected that the charged particles in the active regions
propagating along the jet are accelerated to very high energies. 

Optical flux variations in blazars are often accompanied by colour variations.
Colour variability trends in blazars is still a matter of debate.
Some authors have detected a bluer-when-brighter trend (e.g. Ghosh et al.\ 2000; Raiteri et al.\ 2001;
Villata et al.\ 2002; Gu et al.\ 2006; Agarwal \& Gupta 2015; and references therein) while some others have claimed the opposite i.e. redder-when-brighter
trend (e.g. Raiteri et al.\ 2007; Gaur, Gupta \& Wiita 2012).
Densely sampled, high precision and simultaneous multi frequency data will be helpful to clearly know the CM relationship in blazars on short term basis and will also help in constraining blazar
variability models.

We also studied SED changes associated with our source, corresponding to the observed flux variations along four optical bands.
Based on the location of low energy peak in blazar SED, we have low energy peaked blazars (LBLs), intermediate-energy peaked blazars (IBLs) and
high-energy peaked blazars (HBLs). 
For HBLs, the synchrotron component peaks in the X-ray range, for IBLs it lies in the optical-UV range, and for LBLs in the infrared region.
FSRQs are found to be exclusively low-energy peaked with synchrotron peak located close to optical wavelengths,
thus optical variability studies assist in constraining various theoretical models. They also provide information on the emitting region of
relativistic electrons, as SED changes are most likely caused by variations in the spectra of emitting electrons, which are further caused from differences in the physical parameters of relativistic jets.
The optical synchrotron spectra of blazars follow single power law: F$_\nu$ $\propto$ $\nu^{-\alpha}$, with $\alpha$ being the optical spectral index.

We used our quasi-simultaneous $B$, $V$, $R$, and $I$ data sets and generated 17 optical SEDs for the source spanning 2018 Feb to 2018 July.
We de-reddened the calibrated magnitudes of 3C 279 by subtracting Galactic absorption $A_{B} = 0.104$~mag, $A_{V} = 0.078$~mag, $A_{R} = 0.062$~mag,
and $A_{I} =  0.043$~mag (Cardelli, Clayton, \& Mathis 1989).
Figure 7 displays quasi-simultaneous narrow band optical SEDs of 3C 279, corresponding to 17 different epochs during the observation span.
The faintest fluxes for our target were measured on 2018 May 13 and the maxima was observed on 2018 March 28, while significant variations were detected on other days as evident from the figure.
Due to lack of multi-wavelength observations, we were not able to explore SED changes further.
Multi-wavelength SEDs can also provide vast amount of information about physical parameters of the emitting region.

\begin{figure}
\centering
\epsfig{figure=  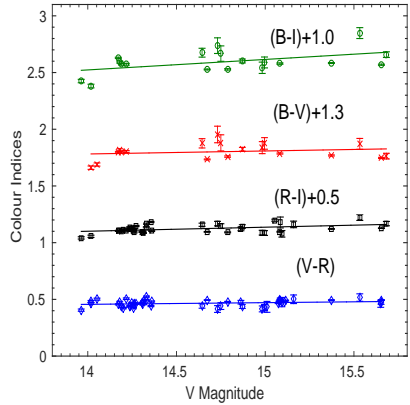,height=2.263in,width=2.401in,angle=0}
  \caption{Colour--magnitude plots on short timescales for 3C 279. The $V$ magnitudes are given on the X-axis and various colour indices are plotted against them.}
\end{figure}

\section{Discussion}

Blazars, being at cosmological distances, cannot be completely resolved using present observational techniques.
Even the high resolution radio telescopes are still not capable of resolving the jet formation region close to the central engine.
Thus investigating variability on diverse timescales will help us
to better understand these objects. Various intrinsic and extrinsic models have been reported to account for temporal variability in blazars.
The intrinsic origin of variability could be either due to relativistic jet or AD instabilities. While the extrinsic ones involve interstellar scintillation (ISS; dominant at low-frequencies),
geometrical effects (Gopal-Krishna \& Wiita 1992) occurring within the jet and gravitational microlensing (Chang \& Refsdal 1979; Bignall et.al. 2003).
Since blazar emission is generally dominated by Doppler boosted non thermal radiation from the relativistic jet, variability on wide range of timescales can be explained by the jet based models.
Flaring activities in blazars on intraday and short timescales could arise due to the emergence and propagation of a new shock which could be due to variations in
the velocity, electron density or magnetic field of the Doppler boosted relativistic jet.
Moreover, optical IDV/STV of blazars can be explained by various models involving irregularities in the jet flow due to ongoing shock; turbulence behind the shock; variations in the outflow parameters due to
magnetic reconnection and turbulence (e.g. Marscher 2014; Calafut \& Wiita 2015; Sironi, Petropoulou, \& Giannios 2015}.

\begin{table}
\caption{Colour-magnitude dependencies and colour-magnitude correlation coefficients on short timescales.}
\textwidth=7.0in
\textheight=10.0in
\vspace*{0.2in}
\noindent

\begin{tabular}{ccccc} \hline 

Colour Indices     &  $m$  &  $c$  &   $r$  & $p$    \\ \hline
($B-I$)          &  0.092 & 1.228 & 0.301  & 0.210  \\     
($B-V$)          &  0.027 & 1.409 & 0.139 & 0.571  \\ 
($R-I$)          &  0.036 & 0.600 & 0.362 & 0.024 \\
($V-R$)          &  0.015 & 0.248 & 0.301 & 0.042 \\
\hline
\end{tabular}\\
$m =$ slope and $c =$ intercept of CI against $V$; \\
$r =$ Spearman coefficient; $p =$ null hypothesis probability \\
\end{table}

\begin{figure}
\hspace{-0.2in}
\epsfig{figure= 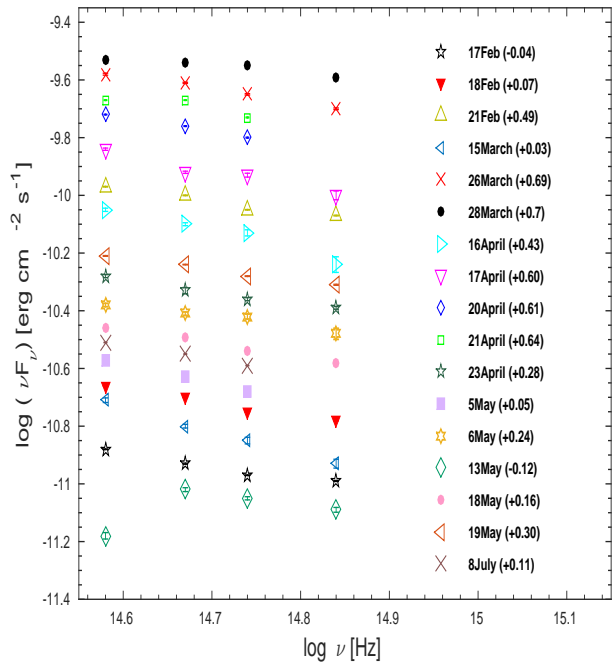,height=3.8in,width=3.6in,angle=0}
\caption{SED results for 3C\,279 in optical frequency range. Different symbols are used for each epoch. The offset used to clearly display SED plot for each date is also mentioned along the date
as shown in the figure.}
\end{figure}

We found a mild BWB trend during our observation span.
BWB trend in blazars can be explained by the presence of two components, a variable one having a flatter slope ($\alpha$; f$_{\nu}$ $\propto$ $\nu^{-\alpha}$) and another stable one with
$\alpha_\mathrm{const} > \alpha$ contributing to overall emission at optical
wavelengths. The variable component contributes to the chromatic behaviour of the source.
Alternatively, the observed BWB trend
can be explained in terms of one component synchrotron model (Fiorucci,
Ciprini, \& Tosti 2004) according to which the emission is boosted to higher frequencies as the energy
release is intense. Studying the behaviour in different optical bands can help us estimate the type of origin for the variability.
Increasing variability amplitude with frequency can be mostly explained by the accelerated electrons at the shock front which then loose energy while
propagating away from the front.
Owing to synchrotron cooling, the high frequency electrons loose energy faster than the low frequency ones.
Due to the closeness of energy bands in the optical regime, a flare should be initiated simultaneously at all optical wavebands, thus observations
on short timescales could miss the detection of the time-lags among various optical bands.
The above emission behaviour of the source was found to be consistent in all four passbands, thus suggesting that $B$, $V$, $R$, and $I$ emissions come from
the same region and by the same physical process.
The present optical data sets for the source can be correlated with
the observations at other wavelengths to investigate its behaviour over
the entire EM
spectrum.

\section{Conclusions}

In this paper, we presented our optical monitoring work which targeted the
blazar, 3C\,279,
observed with seven different telescopes between 2018 February and 2018 July. This helped in studying the intraday to long term flux and colour variability of this source
to understand its nature in the optical regime. We gathered about 716 optical $B$, $V$, $R$, and $I$ frames for 3C\,279.
The analysis of the data revealed that the source was active during the entire monitoring period.

(i) 3C 279 was found to be variable on 5 out of 7 nights on intraday timescales.
The source displayed intraday variability amplitudes ranging from 5.20\% to 17.90\% between 2018 Feb to July.

(ii) All optical passbands ($B$, $V$, $R$, and $I$) clearly displayed flux variability on short timescales while colour variability was found to be very weak.

(iii) The short term variability amplitude was found to increase with frequency, similar to the results of Papadakis et al.\ (2003).
The maximum variability amplitude was found in the $B$ passband of around 177 per cent.

(iv) The source depicted a small flare peaking around 2018 March 28 and later
attained the faintest state around 2018 May 13, while was
found to again brighten afterwards. Unfortunately, due to several observational constraints we could not
obtain data samples to cover this brightening phase of 3C 279.

(v) The above emission behaviours of the source were found to be consistent in all four passbands thus suggesting that $B$, $V$, $R$, and $I$ emissions come from
the same region and by the same physical process which is expected due to the closeness of the bands.

(vi) Using SF analysis we found shortest variability timescale of 23 minutes while the highest reaching upto 115 minutes. Our claim of the variability event on 2019 April 18, over a
timescale of $\sim$ 23 minutes has a low amplitude of $\sim$ 17.90\% (Cellone, Romero, \& Araudo 2007). Also, we have accounted for error sources involved in aperture photometry followed by
quantitative analysis of any detected variations thus we can firmly establish that we have detected a real variability timescale.

(vii) The shortest variability timescale of 23 minutes gives a lower limit on the size of emission region of about $0.31 \times 10^{15}$~cm.
 
(viii) We also studied correlation between the colour indices of the target and its $V$ band magnitude and found that BWB trend was dominant on short timescales.

(ix) Optical spectral indices ranged from 2.3 to 3.0 but displayed no clear trend with time during our observation span. We generated 17 optical SEDs using quasi-simultaneous $B$, $V$, $R$, and $I$
observation points and found
the faintest SED on 2018 May 13 while the brightest was observed on 2018 March 28. Significant variations were observed between the brightest and the faintest SED.

To further understand the variability on diverse timescales for our target 3C\,279 in the optical window, densely sampled observations are
encouraged.
Above optical data sets for the source can be correlated with the
observations at other wavelengths to investigate its behaviour over
the entire EM spectrum. 
In addition to that, simultaneous multi-frequency data will help
us model various emission mechanisms and constrain the theoretical models.

\section*{Acknowledgments}

AA is supported by Science and Engineering Research Board, grant no. PDF/2016/002648.
AA would like to sincerely thank Vipul Agrawal for useful discussions.
We thank the staff of IAO, Hanle and CREST, Hosakote, that made these observations possible. The facilities at IAO and CREST are operated by the Indian Institute of Astrophysics, Bangalore.
Based on data acquired at Complejo Astron\' omico El Leoncito, operated under agreement
between the Consejo Nacional de Investigaciones Cient\'ificas y T\'ecnicas de la Rep\' ublica
Argentina and the National Universities of La Plata, Córdoba and San Juan.
This research was partially supported by the Bulgarian National Science Fund of the Ministry of Education and Science under grants DN 08-1/2016 and DN 18/13-2017.
We also acknowledge support from the Scientific and Technological Research Council of
Turkey (T$\ddot{U}$B$\dot{I}$TAK,  2218-programme), the Republic of Turkey Ministry of Development (PN: 2016K12137 and 3685) and Istanbul University (PN: FBG-2017-23943).

\label{lastpage}

\begin{thebibliography}{}

\bibitem[\protect\citeauthoryear{Abdo et al.}{2010}]{2010ApJ...722..520A} Abdo A. A., et al. 2010, ApJ, 722, 520

\bibitem[\protect\citeauthoryear{Abramowicz \& Nobili}{1982}]{1982Natur.300..506A} Abramowicz M.~A., Nobili L., 1982, Natur, 300, 506 

\bibitem[\protect\citeauthoryear{Acero et al.}{2015}]{2015ApJS..218...23A} Acero F., et al., 2015, ApJS, 218, 23 

\bibitem[\protect\citeauthoryear{Agarwal 
\& Gupta}{2015}]{2015MNRAS.450..541A} Agarwal A., Gupta A.~C., 2015, MNRAS, 450, 541 

\bibitem[\protect\citeauthoryear{Agarwal et 
al.}{2015}]{2015MNRAS.451.3882A} Agarwal A., et al., 2015, MNRAS, 451, 3882 

\bibitem[\protect\citeauthoryear{Agarwal et al.}{2016}]{2016MNRAS.455..680A} Agarwal A., et al., 2016, MNRAS, 455, 680 

\bibitem[\protect\citeauthoryear{Ackermann et al.}{2016}]{2016ApJ...824L..20A} Ackermann M., et al., 2016, ApJ, 824, L20 

\bibitem[\protect\citeauthoryear{Andruchow et al.}{2003}]{2003A&A...409..857A} Andruchow I., Cellone S.~A., Romero G.~E., Dominici T.~P., Abraham Z., 2003, A\&A, 409, 857 

\bibitem[\protect\citeauthoryear{Banerjee et al.}{2019}]{2019MNRAS.487..845B}
Banerjee B., Joshi M., Majumdar P., Williamson K. E., Jorstad S. G., Marscher A. P., 2019, MNRAS, 487, 845

\bibitem[\protect\citeauthoryear{Begelman, Blandford, \& Rees}{1984}]{1984RvMP...56..255B} Begelman M.~C., Blandford R.~D., Rees M.~J., 1984, RvMP, 56, 255 

\bibitem[\protect\citeauthoryear{Bessell, Castelli, 
\& Plez}{1998}]{1998A&A...333..231B} Bessell M.~S., Castelli F., Plez B., 1998, A\&A, 333, 231 

\bibitem[\protect\citeauthoryear{Bhatta et al.}{2013}]{2013A&A...558A..92B} Bhatta G., et al., 2013, A\&A, 558, A92 

\bibitem[\protect\citeauthoryear{Bignall et al.}{2003}]{2003ApJ...585..653B} Bignall H.~E., et al., 2003, ApJ, 585, 653 

\bibitem[\protect\citeauthoryear{Blandford 
\& Rees}{1978}]{1978PhyS...17..265B} Blandford R.~D., Rees M.~J., 1978, PhyS, 17, 265 

\bibitem[\protect\citeauthoryear{Blandford \& K{\"o}nigl}{1979}]{1979ApJ...232...34B} Blandford R.~D., K{\"o}nigl A., 1979, ApJ, 232, 34 

\bibitem[\protect\citeauthoryear{Blinov et al.}{2015}]{2015MNRAS.453.1669B} Blinov D., et al., 2015, MNRAS, 453, 1669 

\bibitem[\protect\citeauthoryear{Bonning et al.}{2012}]{2012ApJ...756...13B} Bonning E., et al., 2012, ApJ, 756, 13 

\bibitem[\protect\citeauthoryear{B{\"o}ttcher et 
al.}{2007}]{2007ApJ...670..968B} B{\"o}ttcher M., et al., 2007, ApJ, 670, 
968 

\bibitem[\protect\citeauthoryear{Burbidge 
\& Rosenberg}{1965}]{1965ApJ...142.1673B} Burbidge E.~M., Rosenberg F.~D., 1965, ApJ, 142, 1673 

\bibitem[\protect\citeauthoryear{B{\"o}ttcher et al.}{2013}]{2013ApJ...768...54B} B{\"o}ttcher M., Reimer A., Sweeney K., Prakash A., 2013, ApJ, 768, 54 

\bibitem[\protect\citeauthoryear{Calafut \& Wiita}{2015}]{2015JApA...36..255C} Calafut V., Wiita P.~J., 2015, JApA, 36, 255 

\bibitem[\protect\citeauthoryear{Cardelli, Clayton, 
\& Mathis}{1989}]{1989ApJ...345..245C} Cardelli J.~A., Clayton G.~C., Mathis J.~S., 1989, ApJ, 345, 245 

\bibitem[\protect\citeauthoryear{Cellone, Romero, \& Araudo}{2007}]{2007MNRAS.374..357C} Cellone S.~A., Romero G.~E., Araudo A.~T., 2007, MNRAS, 374, 357 

\bibitem[\protect\citeauthoryear{Chakrabarti 
\& Wiita}{1993}]{1993ApJ...411..602C} Chakrabarti S.~K., Wiita P.~J., 1993, ApJ, 411, 602

\bibitem[\protect\citeauthoryear{Chang \& Refsdal}{1979}]{1979Natur.282..561C} Chang K., Refsdal S., 1979, Natur, 282, 561 

\bibitem[\protect\citeauthoryear{Chatterjee et al.}{2008}]{2008ApJ...689...79C} Chatterjee R., et al., 2008, ApJ, 689, 79 

\bibitem[\protect\citeauthoryear{Chen}{2018}]{2018ApJS..235...39C} Chen L., 2018, ApJS, 235, 39 

\bibitem[\protect\citeauthoryear{Chiaberge \& Ghisellini}{1999}]{1999MNRAS.306..551C} Chiaberge M., Ghisellini G., 1999, MNRAS, 306, 551 

\bibitem[\protect\citeauthoryear{Clements \& Carini}{2001}]{2001AJ....121...90C} Clements S.~D., Carini M.~T., 2001, AJ, 121, 90 

\bibitem[\protect\citeauthoryear{Cohen et al.}{1971}]{1971ApJ...170..207C} Cohen M.~H., Cannon W., Purcell G.~H., Shaffer D.~B., Broderick J.~J., Kellermann K.~I., Jauncey D.~L., 1971, ApJ, 170, 207 

\bibitem[\protect\citeauthoryear{Collmar et al.}{2007}]{2007ESASP.622..207C} Collmar W., et al., 2007, ESASP, 622, 207 

\bibitem[\protect\citeauthoryear{de Diego}{2010}]{2010AJ....139.1269D} de 
Diego J.~A., 2010, AJ, 139, 1269

\bibitem[\protect\citeauthoryear{Edelson \& Krolik}{1988}]{1988ApJ...333..646E} Edelson R.~A., Krolik J.~H., 1988, ApJ, 333, 646 

\bibitem[\protect\citeauthoryear{Fan
\& Lin}{2000}]{2000ApJ...537..101F} Fan J.~H., Lin R.~G., 2000, ApJ, 537, 101  

\bibitem[\protect\citeauthoryear{Fan et al.}{2005}]{2005ChJAA...5..457F} Fan J.-H., Romero G.~E., Wang Y.-X., Zhang J.-S., 2005, ChJAA, 5, 457 

\bibitem[\protect\citeauthoryear{Fiorucci, Ciprini, 
\& Tosti}{2004}]{2004A&A...419...25F} Fiorucci M., Ciprini S., Tosti G., 2004, A\&A, 419, 25 

\bibitem[\protect\citeauthoryear{Gaur et al.}{2010}]{2010ApJ...718..279G} 
Gaur H., Gupta A.~C., Lachowicz P., Wiita P.~J., 2010, ApJ, 718, 279 

\bibitem[\protect\citeauthoryear{Gaur, Gupta, 
\& Wiita}{2012}]{2012AJ....143...23G} Gaur H., Gupta A.~C., Wiita P.~J., 2012, AJ, 143, 23

\bibitem[\protect\citeauthoryear{Ghisellini et al.}{1997}]{1997A&A...327...61G} Ghisellini G., et al., 1997, A\&A, 327, 61 

\bibitem[\protect\citeauthoryear{Ghosh et al.}{2000}]{2000ApJ...537..638G} 
Ghosh K.~K., Ramsey B.~D., Sadun A.~C., Soundararajaperumal S., Wang J., 
2000, ApJ, 537, 638

\bibitem[\protect\citeauthoryear{Giommi et al.}{2012}]{2012MNRAS.420.2899G} Giommi P., Padovani P., Polenta G., Turriziani S., D'Elia V., Piranomonte S., 2012, MNRAS, 420, 2899 

\bibitem[\protect\citeauthoryear{Gopal-Krishna \& Wiita}{1992}]{1992A&A...259..109G} Gopal-Krishna, Wiita P.~J., 1992, A\&A, 259, 109 

\bibitem[\protect\citeauthoryear{Gopal-Krishna et 
al.}{2003}]{2003ApJ...586L..25G} Gopal-Krishna, Stalin C.~S., Sagar R., 
Wiita P.~J., 2003, ApJ, 586, L25 

\bibitem[\protect\citeauthoryear{Gu, Cao, \& Jiang}{2001}]{2001MNRAS.327.1111G} Gu M., Cao X., Jiang D.~R., 2001, MNRAS, 327, 1111 

\bibitem[\protect\citeauthoryear{Gu et 
al.}{2006}]{2006A&A...450...39G} Gu M.~F., Lee C.-U., Pak S., Yim H.~S., Fletcher A.~B., 2006, A\&A, 450, 39

\bibitem[\protect\citeauthoryear{Gupta et 
al.}{2004}]{2004A&A...422..505G} Gupta A.~C., Banerjee D.~P.~K., Ashok N.~M., Joshi U.~C., 2004, A\&A, 422, 505 

\bibitem[\protect\citeauthoryear{Gupta et al.}{2008}]{2008AJ....136.2359G}
Gupta A.~C., et al., 2008, AJ, 136, 2359

\bibitem[\protect\citeauthoryear{Gupta et al.}{2016}]{2016MNRAS.458.1127G} Gupta A.~C., et al., 2016, MNRAS, 458, 1127 

\bibitem[\protect\citeauthoryear{Guetta et al.}{2004}]{2004A&A...421..877G} Guetta D., Ghisellini G., Lazzati D., Celotti A., 2004, A\&A, 421, 877 

\bibitem[\protect\citeauthoryear{H.E.S.S.~Collaboration et 
al.}{2014}]{2014A&A...571A..39H} H.E.S.S.~Collaboration, et al., 2014, A\&A, 571, AA39 

\bibitem[\protect\citeauthoryear{Heidt
\& Wagner}{1996}]{1996A&A...305...42H} Heidt J., Wagner S.~J., 1996, A\&A, 305, 42

\bibitem[\protect\citeauthoryear{Hunger \& Reimer}{2016}]{2016A&A...589A..96H} Hunger L., Reimer A., 2016, A\&A, 589, A96 

\bibitem[\protect\citeauthoryear{Jockers et al.}{2000}]{2000KFNTS...3...13J} Jockers K., et al., 2000, KFNTS, 3, 13 

\bibitem[\protect\citeauthoryear{Jorstad et al.}{2004}]{2004AJ....127.3115J} Jorstad S.~G., Marscher A.~P., Lister M.~L., Stirling A.~M., Cawthorne T.~V., G{\'o}mez J.-L., Gear W.~K., 2004, AJ, 127, 3115 

\bibitem[\protect\citeauthoryear{Jorstad et al.}{2005}]{2005AJ....130.1418J} Jorstad S.~G., et al., 2005, AJ, 130, 1418 

\bibitem[\protect\citeauthoryear{Jorstad et al.}{2017}]{2017ApJ...846...98J} Jorstad S.~G., et al., 2017, ApJ, 846, 98 

\bibitem[\protect\citeauthoryear{Joshi \& B{\"o}ttcher}{2011}]{2011ApJ...727...21J} Joshi M., B{\"o}ttcher M., 2011, ApJ, 727, 21 

\bibitem[\protect\citeauthoryear{Kaur et al.}{2017}]{2017MNRAS.469.2305K} Kaur N., Sameer, Baliyan K.~S., Ganesh S., 2017, MNRAS, 469, 2305 

\bibitem[\protect\citeauthoryear{Kirk et al.}{1998}]{1998A&A...333..452K}

Kirk J. G., Rieger F. M., Mastichiadis A., 1998, A\&A, 333, 452

\bibitem[\protect\citeauthoryear{Konigl}{1981}]{1981ApJ...243..700K} Konigl A., 1981, ApJ, 243, 700 

\bibitem[\protect\citeauthoryear{Lachowicz, Czerny, \& Abramowicz}{2006}]{2006astro.ph..7594L} Lachowicz P., Czerny B., Abramowicz M.~A., 2006, astro, arXiv:astro-ph/0607594 

\bibitem[\protect\citeauthoryear{Lainela et 
al.}{1999}]{1999ApJ...521..561L} Lainela M., et al., 1999, ApJ, 521, 561 

\bibitem[\protect\citeauthoryear{Lindfors et al.}{2006}]{2006A&A...456..895L} Lindfors E.~J., et al., 2006, A\&A, 456, 895 

\bibitem[\protect\citeauthoryear{Liodakis et al.}{2018}]{2018ApJ...866..137L} Liodakis I., Hovatta T., Huppenkothen D., Kiehlmann S., Max-Moerbeck W., Readhead A.~C.~S., 2018, ApJ, 866, 137 

\bibitem[\protect\citeauthoryear{Liu \& Bai}{2015}]{2015AJ....149..191L} Liu H.~T., Bai J.~M., 2015, AJ, 149, 191 

\bibitem[\protect\citeauthoryear{Lynds, Stockton, \& Livingston}{1965}]{1965ApJ...142.1667L} Lynds C.~R., Stockton A.~N., Livingston W.~C., 1965, ApJ, 142, 1667 

\bibitem[\protect\citeauthoryear{MAGIC Collaboration et al.}{2008}]{2008Sci...320.1752M} MAGIC Collaboration, et al., 2008, Sci, 320, 1752 

\bibitem[\protect\citeauthoryear{Man et al.}{2016}]{2016MNRAS.456.3168M}

Man Z., Zhang X., Wu J., Yuan Q., 2016, MNRAS, 456, 3168

\bibitem[\protect\citeauthoryear{Mangalam \& Wiita}{1993}]{1993ApJ...406..420M} 
Mangalam A.~V., Wiita P.~J., 1993, ApJ, 406, 420 

\bibitem[\protect\citeauthoryear{Marscher 
\& Gear}{1985}]{1985ApJ...298..114M} Marscher A.~P., Gear W.~K., 1985, ApJ, 298, 114 

\bibitem[\protect\citeauthoryear{Maraschi et al.}{1994}]{1994ApJ...435L..91M} Maraschi L., et al., 1994, ApJ, 435, L91 

\bibitem[\protect\citeauthoryear{Markwardt}{2009}]{2009ASPC..411..251M} Markwardt C. B., 2009, ADASS XVIII, ASP Conf. Ser., 411, 251

\bibitem[\protect\citeauthoryear{Marscher}{2014}]{2014ApJ...780...87M}
Marscher, A.~P., 2014, ApJ, 780, article id.\ 87

\bibitem[\protect\citeauthoryear{Marscher \& Travis}{1996}]{1996A&AS..120C.537M} Marscher A.~P., Travis J.~P., 1996, A\&AS, 120, 537 

\bibitem[\protect\citeauthoryear{Miller, Carini, \& Goodrich}{1989}]{1989Natur.337..627M} Miller H.~R., Carini M.~T., Goodrich B.~D., 1989, Natur, 337, 627 

\bibitem[\protect\citeauthoryear{Montagni et al.}{2006}]{2006A&A...451..435M}

Montagni F., Maselli A., Massaro E., Nesci R., Sclavi S., Maesano M., 2006, A\&A, 451, 435

\bibitem[\protect\citeauthoryear{M{\"u}cke et al.}{2003}]{2003APh....18..593M} M{\"u}cke A., Protheroe R.~J., Engel R., Rachen J.~P., Stanev T., 2003, APh, 18, 593 

\bibitem[\protect\citeauthoryear{Nilsson et al.}{2009}]{2009A&A...505..601N} Nilsson K., Pursimo T., Villforth C., Lindfors E., Takalo L.~O., 2009, A\&A, 505, 601 

\bibitem[\protect\citeauthoryear{Papadakis et 
al.}{2003}]{2003A&A...397..565P} Papadakis I.~E., Boumis P., Samaritakis V., Papamastorakis J., 2003, A\&A, 397, 565 

\bibitem[\protect\citeauthoryear{Rafle et al.}{2012}]{2012JSARA...7...33R}

Rafle H., Webb J. R., Bhatta, G., 2012, JSARA, 7, 33

\bibitem[\protect\citeauthoryear{Raiteri et 
al.}{2001}]{2001A&A...377..396R} Raiteri C.~M., et al., 2001, A\&A, 377, 396 

\bibitem[\protect\citeauthoryear{Raiteri et 
al.}{2003}]{2003A&A...402..151R} Raiteri C.~M., et al., 2003, A\&A, 402, 151 

\bibitem[\protect\citeauthoryear{Raiteri et 
al.}{2007}]{2007A&A...473..819R} Raiteri C.~M., et al., 2007, A\&A, 473, 819 

\bibitem[\protect\citeauthoryear{Rani et al.}{2010}]{2010MNRAS.404.1992R} 
Rani B., et al., 2010, MNRAS, 404, 1992 

\bibitem[\protect\citeauthoryear{Romero, Cellone, 
\& Combi}{1999}]{1999A&AS..135..477R} Romero G.~E., Cellone S.~A., Combi J.~A., 1999, A\&AS, 135, 477

\bibitem[\protect\citeauthoryear{Shakura \& Sunyaev}{1973}]{1973A&A....24..337S} Shakura N.~I., Sunyaev R.~A., 1973, A\&A, 24, 337 

\bibitem[\protect\citeauthoryear{Sasada et al.}{2010}]{2010PASJ...62..645S} 
Sasada M., et al., 2010, PASJ, 62, 645 

\bibitem[\protect\citeauthoryear{Simonetti, Cordes, \& Heeschen}{1985}]{1985ApJ...296...46S} Simonetti J.~H., Cordes J.~M., Heeschen D.~S., 1985, ApJ, 296, 46 

\bibitem[\protect\citeauthoryear{Sironi, Petropoulou, \& Giannios}{2015}]{2015MNRAS.450..183S} Sironi L., Petropoulou M., Giannios D., 2015, MNRAS, 450, 183 

\bibitem[\protect\citeauthoryear{Spada et al.}{2001}]{2001MNRAS.325.1559S} 
Spada M., Ghisellini G., Lazzati D., Celotti A., 2001, MNRAS, 325, 1559 

\bibitem[\protect\citeauthoryear{Stalin et al.}{2006}]{2006MNRAS.366.1337S} 
Stalin C.~S., Gopal-Krishna, Sagar R., Wiita P.~J., Mohan V., Pandey A.~K., 
2006, MNRAS, 366, 1337 

\bibitem[\protect\citeauthoryear{Stetson}{1987}]{1987PASP...99..191S} 
Stetson P.~B., 1987, PASP, 99, 191

\bibitem[\protect\citeauthoryear{Stetson}{1992}]{1992ASPC...25..297S} 
Stetson P.~B., 1992, ASPC, 25, 297 

\bibitem[\protect\citeauthoryear{Unwin et al.}{1989}]{1989ApJ...340..117U} Unwin S.~C., Cohen M.~H., Biretta J.~A., Hodges M.~W., Zensus J.~A., 1989, ApJ, 340, 117 

\bibitem[\protect\citeauthoryear{Urry 
\& Padovani}{1995}]{1995PASP..107..803U} Urry C.~M., Padovani P., 1995, PASP, 107, 803 

\bibitem[\protect\citeauthoryear{Villata 
\& Raiteri}{1999}]{1999A&A...347...30V} Villata M., Raiteri C.~M., 1999, A\&A, 347, 30 

\bibitem[\protect\citeauthoryear{Villata et 
al.}{2002}]{2002A&A...390..407V} Villata M., et al., 2002, A\&A, 390, 407 

\bibitem[\protect\citeauthoryear{Wagner 
\& Witzel}{1995}]{1995ARA&A..33..163W} Wagner S.~J., Witzel A., 1995, ARA\&A, 33, 163 

\bibitem[\protect\citeauthoryear{Wehrle et al.}{1998}]{1998ApJ...497..178W} Wehrle A.~E., et al., 1998, ApJ, 497, 178 

\bibitem[\protect\citeauthoryear{Wehrle et al.}{2001}]{2001ApJS..133..297W} Wehrle A.~E., Piner B.~G., Unwin S.~C., Zook A.~C., Xu W., Marscher A.~P., Ter{\"a}sranta H., Valtaoja E., 2001, ApJS, 133, 297 

\bibitem[\protect\citeauthoryear{Wierzcholska et 
al.}{2015}]{2015A&A...573A..69W} Wierzcholska A., Ostrowski M., Stawarz {\L}., Wagner S., Hauser M., 2015, A\&A, 573, AA69

\bibitem[\protect\citeauthoryear{Woo \& Urry}{2002}]{2002ApJ...579..530W} Woo J.-H., Urry C.~M., 2002, ApJ, 579, 530 

\bibitem[\protect\citeauthoryear{Xie et al.}{2002}]{2002MNRAS.334..459X} Xie G.~Z., Liang E.~W., Zhou S.~B., Li K.~H., Dai B.~Z., Ma L., 2002, MNRAS, 334, 459 

\bibitem[\protect\citeauthoryear{Xie et al.}{2003}]{2003AJ....126.2108X} Xie G.-Z., Ma L., Liang E.-W., Zhou S.-B., Xie Z.-H., 2003, AJ, 126, 2108 

\bibitem[\protect\citeauthoryear{Xie, Zhou, \& Liang}{2004}]{2004AJ....127...53X} Xie G.~Z., Zhou S.~B., Liang E.~W., 2004, AJ, 127, 53 

\bibitem[\protect\citeauthoryear{Xie et al.}{2005}]{2005PASJ...57..183X} Xie G.-Z., Chen L.-E., Xie Z.-H., Ma L., Zhou S.-B., 2005, PASJ, 57, 183 

\bibitem[\protect\citeauthoryear{Zhang et al.}{2012}]{2012ApJ...752..157Z} Zhang J., Liang E.-W., Zhang S.-N., Bai J.~M., 2012, ApJ, 752, 157 

\bibitem[\protect\citeauthoryear{Zhang et al.}{2016}]{2016Galax...4...25Z}

Zhang X., Wu J., Man Z., 2016, Galax, 4, 25

\bibitem[\protect\citeauthoryear{Zibecchi, Andruchow, Cellone, Carpintero, Romero \& Combi}{2017}]{2017MNRAS.467..340Z} Zibecchi L., Andruchow I., Cellone S.~A., Carpintero D.~D., Romero G.~E., Combi J.~A., 2017, MNRAS, 467, 340

\end{thebibliography}
\end{document}